\documentclass[10pt,english,british,tightenlines,eqsecnum,
floats,aps,amsmath,amssymb,nofootinbib,superscriptaddress,prd,showpacs,showkeys]
              {revtex4}

\usepackage[latin9]{inputenc}
\usepackage{color}
\usepackage{babel}
\usepackage{amsmath}
\usepackage{amssymb}
\usepackage{graphicx}
\usepackage{esint}
\usepackage[unicode=true,pdfusetitle,
 bookmarks=true,bookmarksnumbered=false,bookmarksopen=false,
 breaklinks=false,pdfborder={0 0 1},backref=false,colorlinks=false]
 {hyperref}
\makeatletter
\@ifundefined{textcolor}{}
{%
 \definecolor{BLACK}{gray}{0}
 \definecolor{WHITE}{gray}{1}
 \definecolor{RED}{rgb}{1,0,0}
 \definecolor{GREEN}{rgb}{0,1,0}
 \definecolor{BLUE}{rgb}{0,0,1}
 \definecolor{CYAN}{cmyk}{1,0,0,0}
 \definecolor{MAGENTA}{cmyk}{0,1,0,0}
 \definecolor{YELLOW}{cmyk}{0,0,1,0}
}

\@ifundefined{date}{}{\date{25-Dec-2019}}
\usepackage{babel}

\usepackage{euscript}\usepackage{epsfig}

\usepackage{amsfonts}

\usepackage{fancyhdr}

\makeatother

\usepackage{amsthm}
\theoremstyle{plain}
\newtheorem*{remark*}{Remark}
\usepackage{cleveref}

\begin{document}

\title{Late time cosmic acceleration in modified S\'{a}ez-Ballester theory}

\author{S. M. M. Rasouli}

\email{mrasouli@ubi.pt}

\affiliation{Departamento de F\'{i}sica, Universidade da Beira Interior, Rua Marqu\^{e}s d'Avila
e Bolama, 6200 Covilh\~{a}, Portugal}

\affiliation{Centro de Matem\'{a}tica e Aplica\c{c}\~{o}es (CMA - UBI),
Universidade da Beira Interior, Rua Marqu\^{e}s d'Avila
e Bolama, 6200 Covilh\~{a}, Portugal}

\affiliation{Department of Physics, Qazvin Branch, Islamic Azad University, Qazvin, Iran}
\author{R. Pacheco}

\email{rpacheco@ubi.pt}


\affiliation{Centro de Matem\'{a}tica e Aplica\c{c}\~{o}es (CMA - UBI),
Universidade da Beira Interior, Rua Marqu\^{e}s d'Avila
e Bolama, 6200 Covilh\~{a}, Portugal}

\author{M. Sakellariadou}
\email{mairi.sakellariadou@kcl.ac.uk}

\affiliation{Theoretical Particle Physics and Cosmology Group, Physics Department,
King's College London,\\ University of London, Strand, London WC2R 2LS, UK.}

\author{P. V. Moniz}

\email{pmoniz@ubi.pt}

\affiliation{Departamento de F\'{i}sica, Universidade da Beira Interior, Rua Marqu\^{e}s d'Avila
e Bolama, 6200 Covilh\~{a}, Portugal}

\affiliation{Centro de Matem\'{a}tica e Aplica\c{c}\~{o}es (CMA - UBI),
Universidade da Beira Interior, Rua Marqu\^{e}s d'Avila
e Bolama, 6200 Covilh\~{a}, Portugal}

\begin{abstract}
We establish an extended version of the modified S\'{a}ez-Ballester (SB) scalar-tensor
theory 
in arbitrary dimensions whose energy momentum tensor as well as
potential are pure geometrical quantities. This scenario emerges
by means of two scalar fields (one is present in the SB theory
and the other is associated with the extra dimension) which
widens the scope of the induced-matter theory.
Moreover, it bears a close resemblance to the standard S\'{a}ez-Ballester  scalar-tensor
theory, as well as other alternative
theories to general relativity, whose construction
includes either a minimally or a non-minimally coupled
scalar field.
However, contrary to those theories, in our framework the energy momentum
tensor and the scalar potential are not
added by hand, but instead are dictated from the geometry.
Concerning cosmological applications, our herein contribution
brings a new perspective.
We firstly show that the dark energy sector can be naturally
retrieved within a strictly geometric perspective, and we subsequently analyze it.
Moreover, our framework may
provide a hint to understand the physics of lower gravity theories.

 \end{abstract}

\medskip



\maketitle

\section{Introduction}

A unification of gravitational, electromagnetic
and scalar interactions was proposed through the
Kaluza-Klein theory \cite{Kaluza21,Klein26,OW97}.
Additionally, the main motivation in the induced matter interpretation of the Kaluza-Klein
gravity was  to show that the energy momentum
tensor, which is added phenomenologically to the right hand side of
Einstein equations, would instead have a geometric origin \cite{PW92}.
It has been shown that the classical
tests of general relativity and the experimental limits on
violation of the equivalence principle do not disqualify any higher
dimensional theories of gravity \cite{LO00,O00,OWM07}.
Concerning astrophysical/cosmological applications, such induced
sectors or their extensions can play the role of either ordinary matter, dark matter or dark
energy \cite{OWM07,DRJ09, Ponce1, RJ10, RFS11, RFM14, RM16, SPS18, RM19}.

In order to obtain an accelerating scale factor within general relativity, the right
hand side of the Einstein field equations should bear a contribution whose pressure is negative.
More precisely, the equation of state (EoS)
parameter $W$ (pressure to density ratio) must be less than $-1/3$.
To achieve such a late time dynamics, the cosmological constant and
the quintessence proposal are the most well-known candidates to constitute the dark energy.

However, explaining the manifestation of the
cosmological constant strictly within classical cosmology in
four dimensions might be a difficult procedure \cite{WMO08}.
Therefore, the cosmological constant has also been
investigated by employing higher-dimensional
theories (see for instance \cite{RS83,OW97,KT02,KZZL19}, and references therein).
Within the context of the standard induced-matter theory (IMT) \cite{PW92}, by employing a canonical
metric, which bears resemblance to the synchronous one
of standard four-dimensional cosmology, it has been shown that Einstein equations
with a cosmological constant are indeed generated \cite{WMO08,W11}.

  Furthermore, the simplest model of a quintessence is obtained by
 including a scalar field $\phi$ (which may respond to a corresponding
potential energy $V(\phi)$) into the Einstein-Hilbert action,
 such that it is minimally coupled to the space-time curvature.
 In this model, by admitting the usual assumption $V\geq 0$, it can be
shown that $W$ is bounded,  $-1\leq W\leq 1$ \cite{F02}.
This implies that the canonical scalar field model with a positive
 potential energy does not yield a super-acceleration
 (whose EoS parameter is restricted as $W<-1$).
 If $V$ takes negative values, a negative energy density may
 emerge \cite{NJP15}.

Hence, the objective of this study is twofold.
Firstly, we extend the induced matter
theory to a generalised setting exhibiting the following features:
(i) relying on the SB theory \cite{SB85-original}, the Einstein-Hilbert action comprises a
non-canonical minimally coupled scalar field (see also \cite{SSL09}); (ii)
instead of a five-dimensional space-time, we assume an arbitrary dimensional bulk;
 (iii)
contrary to the standard procedure \cite{PW92}, we employ a general coordinate free framework, which we
subsequently apply to a particular case of the metric taken in the original IMT \cite{PW92}, its
corresponding extended version \cite{RRT95}
and some modified scalar-tensor theories \cite{RFM14,RM18}.
Secondly, we analyze the analytic solutions for an extended Friedmann-Lema\^{i}tre-Robertson-Walker (FLRW)
metric as well as for a canonical metric in arbitrary dimensions.
Specifically, we investigate the conditions under which an
accelerated stage is dominant in the universe.

More precisely, we will show that, by admitting specific conditions
for the integration constants and parameters, the induced matter on the
hypersurface not only may have the properties of ordinary matter but
can also be constructed towards playing the role of dark energy.
Furthermore, the total energy density and
total pressure components have two
sectors: the induced sector and the sector associated with the scalar field $\phi$.
For each subsequent solution, we will investigate the properties
of those fluids and check whether the  weak energy condition is
satisfied or not. Another motivation of assuming arbitrary dimensions
can be to explore feasible relations between the standard four
dimensional gravity and lower gravity
theories,\footnote{Studying such problems will not be within
the scope of our present investigation.} see, e.g., \cite{RRT95}.

This paper is organised as follows.
In  Section \S \ref{Set up}, by taking the SB
scalar-tensor theory, and assuming a general metric
in $(D+1)$-dimensional space-time, we will establish an
extended version of the MSBT \cite{RM18} on a $D$-dimensional hypersurface.
This formalism is constructed by employing
an approach from a purely mathematical point of view
which is different from those previously employed.
In Section \S \ref{Bulk-solutions}, for an extended FLRW metric in
vacuum, we will show that there are two constants of motion
associated with the SB field equations in the $(D+1)$-dimensional bulk.
Then, we will obtain two different classes of analytic solutions: the exponential-law
and power-law solutions for the scale factors.
Their corresponding SB scalar field can take either
exponential, logarithmic or power-law forms.
In Section \S \ref{Red-cosm}, we will analyze the cosmological solutions on the hypersurface.
We will focus on the interesting solutions which not only yield
an accelerated scale factor for the universe but also have a contracting
extra dimension, which can be applicable for the late time epoch.
The last section includes a summary as well as our conclusions.
Moreover, concerning the cosmological constant in noncompact Kaluza-Klein
frameworks, we will provide a complementary
discussion from a canonical gauge perspective.

\section{Modified S\'{a}ez-Ballester scalar-tensor theory in arbitrary dimension}
\label{Set up}
\indent

In the following, we extend the procedure established in \cite{RM18} to arbitrary
dimensions.
In analogy with the four-dimensional action proposed in~\cite{SB85-original}, let us adapt
the $(D+1)$-dimensional generalised counterpart of the same theory as
\begin{equation}\label{SB-5action}
{\cal S}^{^{(D+1)}}=\int d^{D+1}x \sqrt{\Bigl|{}{\Upsilon}^{^{(D+1)}}\Bigr|} \,\left[R^{(D+1)}
-{\cal W}\phi^n\,{\Upsilon}^{ab}\,(\nabla_a\phi)(\nabla_b\phi)+\chi\,
L\!^{^{(D+1)}}_{_{\rm matt}}\right].
\end{equation}
Throughout this paper we use the geometrised units where $G=1=c$ (where $G$ and $c$ are the Newton gravitational
constant and the speed of light in vacuum, respectively) and $\chi=8\pi$.
Therefore, $\phi$ is a dimensionless scalar field
(designated as the SB scalar field)
and $n$, ${\cal W}$ are dimensionless parameters;
${\Upsilon}^{^{(D+1)}}$ stands for the determinant of the $(D+1)$-dimensional
metric ${\Upsilon}_{{ab}}$, whose Ricci scalar was
denoted by $R^{^{(D+1)}}$ and $\nabla_a$ denotes the covariant
derivative in the bulk.
Moreover, the Greek and Latin indices take values from zero
to $D-1$ and to $D$, respectively.

It is worth noting that in establishing the IMT \cite{PW92}, or even the modified scalar-tensor
theories, see, e.g., \cite{Ponce1}, an apparent vacuum\footnote{In our herein paper, as in all types of the
scalar-tensor theories, a ``vacuum''
space-time will be used with the meaning that the ``ordinary matter'' is absent.
Taking a higher dimensional vacuum universe (the Kaluza's first key assumption) was inspired by
Einstein \cite{OW97}.} bulk has been considered.
However, in this work, we would establish a more extended version of a reduced theory by taking
a non-vanishing energy momentum tensor, $L\!^{^{(D+1)}}_{_{\rm matt}}\neq0$.
(One is free to postulate matter fields in higher dimensions, which yields
complicated Kaluza-Klein models. However, such an approach runs the risk of
introducing so many degrees of freedom that the theory is no longer testable.
In this regard, in the next section, we assume a vacuum bulk.)

The field equations corresponding to the action (\ref{SB-5action}) are given by
\begin{equation}\label{(D+1)-equation-1}
G^{^{(D+1)}}_{ab}={\cal W}\phi^{n}\left[(\nabla_a\phi)(\nabla_b\phi)
-\frac{1}{2}{\Upsilon}_{ab}(\nabla^c\phi)(\nabla_c\phi)\right]+\chi\,T^{^{(D+1)}}_{ab}
\end{equation}
and
\begin{equation}\label{(D+1)-equation-2}
2\phi^n\nabla^2\phi
+n\phi^{n-1}(\nabla_a\phi)(\nabla^a\phi)=0,
\end{equation}
where $G^{^{(D+1)}}_{ab}$ is the Einstein tensor of $(D+1)$-dimensional
space-time, $\nabla^2\equiv\nabla_a\nabla^a$ and
$T^{^{(D+1)}}_{ab}$ denotes the energy momentum tensor of any ordinary matter fields in the bulk.
Moreover, $T^{^{(D+1)}}_{ab}$ does not depend on the SB scalar field and obeys a conservation law.
From Eq.~(\ref{(D+1)-equation-1}), we easily obtain
\begin{equation}\label{(D+1)-equation-3}
R^{^{(D+1)}}={\cal W}\phi^{n}(\nabla_a\phi)(\nabla^a\phi)-\frac{2\chi}{D-1} T^{^{(D+1)}},
\end{equation}
where $T^{^{(D+1)}}={\Upsilon}^{ab}T^{^{(D+1)}}_{ab}$.


In order to establish an induced gravity on a $D$-dimensional
hypersurface, in contrast to the procedure used to construct the original IMT and
 modified scalar-tensor theories \cite{PW92,RRT95,RFM14,RM18},
we shall start by extracting, in a rather general and
coordinate free framework, the fundamental equations that provide the
geometrical quantities associated with the hypersurface from those associated with
 the bulk. Subsequently, we consider a particular
 case associated with a specific metric, which is the same
  metric employed in the original IMT \cite{PW92}, its
corresponding extended version \cite{RRT95}
and the recent modified scalar-tensor theories \cite{RFM14,RM18}.
Finally, in Section \S \ref{induced-dynamics}, we will recover the modified $D$-dimensional SB field
equations on a hypersurface.

\subsection{Hypersurfaces of semi-Riemannian manifolds}

The bulk is seen as a semi-Riemannian manifold $\bar\Sigma$ of dimension $D+1$ with metric tensor $dS^2$.
Let $\Sigma \subset  \bar \Sigma$ be a semi-Riemannian hypersurface of sign $\epsilon=\pm1$, that
is, any normal vector $u$ satisfies  $dS^2(u,u)>0$ if $\varepsilon=1$ and  $dS^2(u,u)<0$ if $\varepsilon=-1$.
The induced metric on $\Sigma$ will also be denoted by $dS^2$. Fixing a unit normal vector field $N$, the second
fundamental form $\theta$ of $\Sigma $ is then given by
\begin{equation}\label{sf}
\theta(X,Y)N= \nabla_XY- \mathcal{D}_XY,
\end{equation} where $X,Y$ are tangent vector fields on $\Sigma $, and
$\mathcal{D}$ and  $\nabla$ are the covariant derivatives
of $\Sigma$ and $ \bar \Sigma$, respectively, with respect to $dS^2$.

Consider the \emph{Gauss-Codazzi equations} \cite{Oneil} (see also \cite{Koiso}):
\begin{align}\label{gauss}
 dS^2( {\mathcal{R}}^{(D+1)}_{XV}Y,W)&= dS^2(\mathcal{R}^{(D)}_{XV}Y,W)
 -\epsilon \theta(X,Y)\theta(W,V)+\epsilon \theta(X,W)\theta(Y,V),\\\nonumber\\
\label{codazzi}
dS^2({\mathcal{R}}^{(D+1)}_{XY}Z,N)&=(\mathcal{D}_Y \theta)(X,Z)-(\mathcal{D}_X \theta)(Y,Z),
  \end{align}
where $\mathcal{R}^{(D)}$ and ${\mathcal{R}}^{(D+1)}$ are the curvature (1,3) tensors of $\Sigma$ and
$ \bar \Sigma$, respectively; for any
tangent vector fields $X_1,X_2,X_3$ of $\Sigma$, $$(\mathcal{D}_{X_1}
\theta)(X_2,X_3)=X_1\theta(X_2,X_3)- \theta(\mathcal{D}_{X_1}X_2,X_3)-\theta(X_2,\mathcal{D}_{X_1}X_3).$$
Set
$$\vartheta(X,Y)=dS^2( \mathcal{R}^{(D+1)}_{XN}Y,N).$$
From \eqref{gauss} and \eqref{codazzi} we obtain the following
formulae for the Ricci tensor ${\mathrm{Ric}}^{(D+1)}$ of $\bar \Sigma$:
\begin{align}\label{XY}
{\mathrm{Ric}}^{(D+1)}(X,Y)&=\mathrm{Ric}^{(D)}(X,Y)+\epsilon \vartheta(X,Y)
-\epsilon\theta(X,Y)\mathcal{H} +\epsilon  \theta^2(X,Y),\\\label{XN}
{\mathrm{Ric}}^{(D+1)}(X,N)&=\delta P(X),\\
{\mathrm{Ric}}^{(D+1)}(N,N)&=\mathrm{tr}\,\vartheta,\label{NN}
\end{align}
where  $\mathrm{Ric}^{(D)}$ is the Ricci tensor of $\Sigma$,
$\mathcal{H}=\mathrm{tr}\, \theta $ is the mean curvature, $\delta$
is the divergence operator, $P$ is the tensor defined by
\begin{equation}
P=\mathcal{H}dS^2-\theta,
\end{equation}
 and $\theta^2$ is defined by $\theta^2(X,Y)=\mathrm{tr}\,\theta(X,\cdot)\theta(Y,\cdot)$.
 The Ricci scalar curvatures $ R^{(D)}$ and $ R^{(D+1)}$ of
 $\Sigma$ and $\bar \Sigma$, respectively, are then related by
\begin{equation}\label{scalarcurvatures}
 R^{(D+1)} =R^{(D)}+2\epsilon\, \mathrm{tr}\,\vartheta -\epsilon(\mathcal{H}^2-l^2),
\end{equation}
where $l^2$ is the square of the length of $\theta$.
We conclude that the Einstein tensors $ G^{(D)}$ and $G^{(D+1)} $ satisfy:
\begin{align}\label{GXY}
G^{(D+1)}(X,Y)&=G^{(D)}(X,Y)+\epsilon \vartheta(X,Y)-\epsilon\theta(X,Y)\mathcal{H} +\epsilon \theta^2(X,Y) -\frac{\epsilon}{2}\big(2\,\mathrm{tr}\,\vartheta-\mathcal{H}^2+l^2 \big)dS^2(X,Y),\\
\label{GXN}G^{(D+1)}(X,N)&=\delta P(X),\\
G^{(D+1)}(N,N)&=\mathrm{tr}\, \vartheta-\frac{\epsilon}{2}\left[R^{(D)}+2\epsilon\,
\mathrm{tr}\,\vartheta -\epsilon(\mathcal{H}^2-l^2) \right].
\end{align}
Let us now fix the local coordinates  $x^0,x^1,\ldots, x^{D-1},x^D\equiv y$ of $\bar\Sigma$,
such that the hypersurface $\Sigma$ is given by $y=y_0$ for some constant $y_0$.
We assume that the metric $dS^2$ is of the form\footnote{We should emphasize that as we do not impose
the cylinder condition, without losing the algebraic generality, we have set the electromagnetic
potentials equal to zero \cite{OW97}.}
\begin{equation}\label{global-metric}dS^{2}=\Upsilon_{ab}(x^c)dx^{a}dx^{b}=
g_{\alpha\beta}(x^\mu,y)dx^{\alpha}dx^{\beta}+
\epsilon\psi^2\left(x^\mu,y\right)dy^{2},
\end{equation}
where
$\psi$ is a smooth function on $\bar \Sigma$.

For any function $\varphi$ on $\bar \Sigma$, denote $\overset{*}\varphi\equiv
\frac{\partial \varphi}{\partial y}$. Fix the unit normal vector field $N=\frac{\epsilon}{\psi}\frac{\partial}{\partial y}$.
In terms of these coordinates, the tensors above can be written as follows:
\begin{align}
 \label{sfc} \theta_{\alpha\beta}&=-\frac{1}{2\psi} \overset{*}g_{\alpha\beta},\\\label{theta2}
  \theta^2_{\alpha\beta}&=\frac{1}{4\psi^2}  g^{\mu\nu}\overset{*}g_{\mu\alpha}\overset{*}g_{\nu\beta},\\
\vartheta_{\alpha\beta}&=-\frac{\epsilon}{\psi}(\mathcal{D}d\psi)_{\alpha\beta}+\frac{1}{2\psi^2}
\left(\frac{\overset{*}\psi}{\psi}\overset{*}g_{\alpha\beta} -\overset{**}{g}_{\alpha\beta}+\frac12   g^{\mu\nu}\overset{*}g_{\mu\alpha}\overset{*}g_{\nu\beta}    \right),\\\label{P-mono}
P_{\alpha\beta}&=\frac{1}{2\psi}\big(\overset{*}g_{\alpha\beta}-g_{\alpha\beta}g^{\mu\nu}\overset{*}g_{\mu\nu}\big),
\\\label{trvartheta}
\mathrm{tr}\,\vartheta &=-\frac{\epsilon}{\psi}\mathcal{D}^2\psi
+\frac{1}{2\psi^2}\left(\frac{\overset{*}\psi}{\psi}g^{\mu\nu}
\overset{*}g_{\mu\nu} -g^{\mu\nu}\overset{**}{g}_{\mu\nu}+\frac12
g^{\mu\nu}g^{\lambda\sigma}\overset{*}g_{\mu\lambda}\overset{*}g_{\nu\sigma}  \right),
\\
  \mathcal{H}&=-\frac{1}{2\psi}  g^{\mu\nu}\overset{*}g_{\mu\nu},\\\label{l2}
  l^2&=\frac{1}{4\psi^2}  g^{\mu\nu}g^{\lambda\sigma}\overset{*}g_{\mu\lambda}\overset{*}
  g_{\nu\sigma}=-\frac{1}{4\psi^2}\overset{*}g^{\mu\sigma}\overset{*}g_{\mu\sigma},
\end{align}
where $\mathcal{D}^2\psi=\mathrm{tr}\mathcal{D}d\psi=\mathcal{D}^\alpha\mathcal{D}_\alpha\psi$
is the Laplacian of $\psi$ on $\Sigma$ and $(\mathcal{D}d\psi)_{\alpha\beta}=\mathcal{D}_\alpha\mathcal{D}_\beta\psi$.
\\
Combining  Eq.~\eqref{GXY} with Eqs.~\eqref{sfc}~--~\eqref{l2}, we obtain
\begin{eqnarray}\nonumber
  G_{\alpha\beta}^{^{(D+1)}} &=& G_{\alpha\beta}^{^{(D)}}-\frac{{\cal D}_\alpha{\cal D}_\beta\psi}{\psi}+ \frac{g_{\alpha\beta}\mathcal{D}^2\psi}{\psi}
+\frac{\epsilon}{2\psi^{2}}\left(\frac{{\overset{*}\psi}}{\psi}\,{\overset{*}g}_{\alpha\beta}-{\overset{**}g}_{\alpha\beta}
+g^{\mu\nu}\overset{*}g_{\mu\alpha}\overset{*}g_{\nu\beta}-\frac{1}{2}
g^{\mu\nu}\overset{*}{g}_{\mu\nu}{\overset{*}g}_{\alpha\beta}\right)\\&+&\frac{\epsilon g_{\alpha\beta}}{8\psi^2}
 \left[\overset{*}{g}^{\mu\nu}\overset{*}{g}_{\mu\nu}
+\left(g^{\mu\nu}{\overset{*}g}_{\mu\nu}\right)^{2}\right]  - \frac{\epsilon g_{\alpha\beta} }{2\psi^2}\left(\frac{\overset{*}\psi}{\psi}g^{\mu\nu}\overset{*}g_{\mu\nu} -g^{\mu\nu}\overset{**}{g}_{\mu\nu}+\frac12   g^{\mu\nu}g^{\lambda\sigma}\overset{*}g_{\mu\lambda}\overset{*}g_{\nu\sigma}\right).    \label{dEinstein}
\end{eqnarray}

\begin{remark*} \emph{Equations \eqref{sfc}~--~\eqref{l2}  admit a particular simple form under some natural geometrical assumptions on the hypersurface $\Sigma$ with respect to the metric \eqref{global-metric}. For example:
\begin{enumerate}\item
Recall that $\Sigma$ is a \emph{totally geodesic} hypersurface if the geodesics of $\Sigma$ are geodesics of the bulk $\bar \Sigma$. This condition is equivalent to the identically vanishing of the second fundamental form $\theta$, that is, $\overset{*}g_{\alpha\beta}=0$ on $\Sigma$.
For example, the FLRW metric \eqref{DO-metric} satisfies this condition.
\item A weaker condition is that of minimality.  Recall that  \emph{minimal}  hypersurface is a hypersurface that locally minimises its volume. This condition is equivalent to the identically vanishing of the mean curvature $\mathcal{H}$. It follows that $\Sigma$ is  minimal if and only if
     $g^{\alpha\beta}\overset{*}g_{\alpha\beta}=0$ on $\Sigma$.
%
\item  For each fixed value of $y$ we have a hypersurface $\Sigma_y$, with $\Sigma_{y_0}=\Sigma$. This gives a foliation $\mathcal{F}=\{\Sigma_{y}\}$  on $\bar{\Sigma}$. In general, a foliation is called a \emph{Riemannian foliation} if any geodesic normal to a leaf remains normal to all leaves it intersects. With respect to the foliation $\mathcal{F}$, it is easy to see that this is equivalent to the condition
    $\mathcal{D}_\mu\psi=0$.
\end{enumerate}}
\end{remark*}
\begin{remark*}
  \emph{If the bulk $\bar \Sigma$ is Ricci flat and the metric \eqref{global-metric} is a vacuum solution, then Eq.~\eqref{NN} implies $\mathrm{tr}\,\vartheta=0$, and we see that, since
$G_{\mu\nu}^{^{(D+1)}}=0$, Eq.~\eqref{dEinstein} coincides with Eq.~(17) in \cite{PW92}.}
\end{remark*}

The second fundamental form \eqref{sf} relates the covariant differentiations in $\Sigma$ and $\bar \Sigma$. Hence,
given a function $\phi$ on $\bar\Sigma$, we can use \eqref{sf} and \eqref{sfc}, to get (see also \cite{Ponce2}):
 \begin{eqnarray}\label{rel.1}
(\nabla^a\phi)(\nabla_a\phi)\!\!&=&\!\!({\cal D}^\alpha\phi)({\cal D}_\alpha\phi)+
\epsilon\Big(\frac{\overset{*}\phi}{\psi}\Big)^2,\\
\label{rel.2}
\nabla_\alpha\nabla_\beta\phi\!\!&=&\!\!{\cal D}_\alpha{\cal D}_\beta\phi+
\frac{\epsilon}{2\psi^2} \overset{*}g_{\alpha\beta}\overset{*}\phi,\\
\label{rel.3} \nabla^2\phi\!\!&=&\!\!{\cal D}^2\phi+\frac{({\cal
D}_\mu\psi)({\cal D}^\mu\phi)}{\psi}
+\frac{\epsilon}{\psi^2}\left[{\overset{**}\phi}+\overset{*}\phi
\left(\frac{g^{\mu\nu}\overset{*}g_{\mu\nu}}{2}-\frac{{\overset{*}\psi}}{\psi}\right)\right],\\
\nabla_{_{D}}\nabla_{_{D}}\phi\!\!&=&\!\!\epsilon\psi({\cal D}_\mu\psi)({\cal D}^\mu\phi)
+{\overset{**}\phi}-\frac{\overset{*}\psi\,\overset{*}\phi}{\psi}.
\label{rel.4}
\end{eqnarray}

\subsection{The effective field equations}\label{induced-dynamics}

In order to construct the dynamics on the hypersurface, let us divide the procedure in four stages as follows.

\begin{enumerate}
  \item
We will obtain a dynamical equation for the scalar
field $\psi$.
Letting $a\rightarrow D$ and $b\rightarrow D$ in
Eq.~(\ref{(D+1)-equation-1}) and using Eq.~(\ref{(D+1)-equation-3}), we get
\begin{eqnarray}\label{Rdd-1}
R^{^{(D+1)}}_{DD}=\left(\frac{\epsilon \chi \psi^2}{1-D}\right)T^{^{(D+1)}}+\chi T_{DD}^{^{(d+1)}}+{\cal W}\phi^{n}\left(\overset{*}\phi\right)^{2},
\end{eqnarray}
where $R_{ab}^{(D+1)}$ is the Ricci tensor in the bulk.
Combining with Eqs.~\eqref{NN} and \eqref{trvartheta}, we obtain
\begin{eqnarray}\label{D2say}
\frac{{\cal D}^2\psi}{\psi}=-\frac{\epsilon}{2\psi^2}
\left[g^{\lambda\beta}\overset{**}{g}_{\lambda\beta}
+\frac{1}{2}\overset{*}{g}^{\lambda\beta}\,\overset{*}{g}_{\lambda\beta}
-\frac{g^{\lambda\beta}\overset{*}{g}_{\lambda\beta}\overset{*}{\psi}}{\psi}\right]
-\frac{\epsilon{\cal W}\phi^n({\overset{*}\phi})^2}{\psi^2}
+\chi\left[\frac{T^{^{(D+1)}}}{D-1}-\frac{\epsilon T^{^{(D+1)}}_{_{DD}}}{\psi^2}\right],
\end{eqnarray}
which is one of the effective field equations associated with the MSBT.

\item
 Letting $a\rightarrow\mu$ and $b\rightarrow\nu$ in
Eq.~(\ref{(D+1)-equation-1}), we
obtain the $D$-dimensional part of $(D+1)$-dimensional counterpart as
\begin{eqnarray}\label{d+1-Einstein}
G_{\mu\nu}^{^{(D+1)}}\!\!\!&=&{\cal W}\phi^{n}
\left[({\cal D}_\mu\phi)({\cal D}_\nu\phi)-\frac{1}{2}
g_{\mu\nu}({\cal D}_\beta\phi)({\cal D}^\beta\phi)\right]
-\frac{\epsilon{\cal W}g_{\mu\nu}}{2}\left(\frac{\overset{*}\phi}{\psi}\right)^2\phi^n+\chi T^{^{(D+1)}}_{\mu\nu}.
\end{eqnarray}
Using Eq.~\eqref{dEinstein} we obtain
\begin{eqnarray}\label{BD-Eq-DD}
G_{\mu\nu}^{^{(D)}}=\chi T_{\mu\nu}^{^{(D)[{\rm eff}]}} +
{\cal W}\phi^n\left[({\cal D}_\mu\phi)({\cal D}_\nu\phi)-
\frac{1}{2}g_{\mu\nu}({\cal D}_\alpha\phi)({\cal
D}^\alpha\phi)\right]-\frac{1}{2}g_{\mu\nu}V(\phi),
\end{eqnarray}
where $V(\phi)$ denotes an induced scalar potential (in stage (3), we will show
how it is related to the other quantities of the model).
$T_{\mu\nu}^{^{(D)[{\rm eff}]}}$ is defined as follows:
\begin{eqnarray}
\label{T1}  T_{\mu\nu}^{^{(D)[{\rm eff}]}}&\equiv&  T_{\mu\nu}^{^{(D+1)}}+T_{\mu\nu}^{^{[\rm SB]}},\hspace{10mm}{\rm where}\\\nonumber\\
   \label{matt.def}
\label{IMTmatt.def}
\chi T_{\mu\nu}^{^{[\rm SB]}}&\equiv &
\frac{{\cal D}_\mu{\cal D}_\nu\psi}{\psi}-\frac{g_{\mu\nu}\mathcal{D}^2\psi}{\psi}
-\frac{\epsilon}{2\psi^{2}}\left(\frac{{\overset{*}\psi}{\overset{*}g}_{\mu\nu}}{\psi}-{\overset{**}g}_{\mu\nu}
+g^{\lambda\alpha}\overset{*}{g}_{\mu\lambda}{\overset{*}g}_{\nu\alpha}
-\frac{1}{2}g^{\alpha\beta}\overset{*}{g}_{\alpha\beta}{\overset{*}g}_{\mu\nu}\right)+\frac{1}{2}g_{\mu\nu}V(\phi)\cr\cr\cr
 \!\!\!&&\!\!+\frac{\epsilon{g}_{\mu\nu}}{2\psi^2}
\left(\frac{\overset{*}\psi}{\psi}g^{\alpha\beta}\overset{*}g_{\alpha\beta} -g^{\alpha\beta}\overset{**}{g}_{\alpha\beta}+\frac12   g^{\alpha\beta}g^{\lambda\sigma}\overset{*}g_{\alpha\lambda}\overset{*}g_{\beta\sigma}   \right )-\frac{\epsilon g_{\mu\nu}}{8\psi^2}
\left[{\overset{*}g}^{\alpha\beta}{\overset{*}g}_{\alpha\beta}
+\left(g^{\alpha\beta}{\overset{*}g}_{\alpha\beta}\right)^{2}\right].
\end{eqnarray}

\item
 We will derive the wave equation correspond to Eq.~(\ref{(D+1)-equation-2}) on
a $D$-dimensional hypersurface. In this respect,  we use Eqs.~\ref{rel.1})
and (\ref{rel.3}) into Eq.~(\ref{(D+1)-equation-2}), to get
\begin{eqnarray}\label{D2-phi}
2\phi^n{\cal D}^2\phi+n\phi^{n-1}({\cal D}_\alpha\phi)({\cal D}^\alpha\phi)
-\frac{1}{\cal W}\frac{dV(\phi)}{d\phi}=0,
\end{eqnarray}
where
\begin{eqnarray}\label{v-def}
 \frac{dV(\phi)}{d\phi}\equiv-\frac{2{\cal W}\phi^n}{\psi^2}
\Bigg\{\psi({\cal D}_\alpha\psi)({\cal D}^\alpha\phi)
+\frac{n\epsilon}{2\phi}\Big(\overset{*}\phi\Big)^2
+\epsilon\Big[{\overset{**}\phi}+{\overset{*}\phi}
\Big(\frac{g^{\mu\nu}{\overset{*}g}_{\mu\nu}}{2}-\frac{{\overset{*}\psi}}{\psi}\Big)\Big]\Bigg\}.
\end{eqnarray}

\item
We will obtain the fourth effective field equation associated with the MSBT.
 Letting $a\rightarrow\alpha$ and $b\rightarrow D$ in Eq.~(\ref{(D+1)-equation-1}) and
 employing \eqref{XN} and \eqref{GXN} associated with the metric~(\ref{global-metric}), we retrieve
 \begin{eqnarray}\label{P-Dynamic}
G_{\alpha D}^{^{(D+1)}}=R_{\alpha D}^{^{(D+1)}}=\psi P^{\beta}{}_{\alpha;\beta}&=&\!\!
\chi T^{^{(D+1)}}_{\alpha D}+{\cal W}\phi^n{\overset{*}\phi}({\cal
D}_\alpha\phi),
\end{eqnarray}
where $P_{\alpha\beta}$ is given by Eq.~\eqref{P-mono} and $\delta P=P^{\beta}{}_{\alpha;\beta}dx^\alpha$.

\end{enumerate}


\begin{remark*}\emph{It is worth mentioning a few comments regarding the MSBT framework we have adopted.
\begin{enumerate}
\item The set of effective field Eqs.~(\ref{D2say}), (\ref{BD-Eq-DD}), (\ref{D2-phi})
and (\ref{P-Dynamic}) in the particular case where $D=4$ reduce
to those obtained on a four-dimensional hypersurface, see \cite{RM18}.
However, in the present study we have adopted a reduction procedure which
is different from the one employed in \cite{RM18}. As a matter of fact, Eq.~\eqref{dEinstein}
gives a pure geometrical relation between the Einstein
tensors $G^{(D+1)}$ and $G^{(D)}$ that unravels the term $\mathrm{tr}\,\vartheta$
given by Eq.~\eqref{trvartheta}. As we have observed above, this term vanishes if
the bulk is Ricci flat with respect to the metric $dS^2$, otherwise
it must be included in the term $T_{\mu\nu}^{^{[\rm IMT]}}$ (see Eq.~\eqref{IMTmatt.def}),
which is the contribution from the geometry of the extra dimension to
$T_{\mu\nu}^{^{(D)[{\rm eff}]}}$.
\item
We can easily show that  Eqs.~(\ref{BD-Eq-DD}) and
(\ref{D2-phi}) can be derived from the
action
\begin{equation}\label{induced-action}
 {\cal S}^{^{(D)}}=\int d^{^{\,D}}\!x \sqrt{-g}\,\left[R^{^{(D)}}-{\cal W}\phi^n\, g^{\alpha\beta}\,({\cal
D}_\alpha\phi)({\cal D}_\beta\phi)-V(\phi)+\chi\,
L\!^{^{(D)}}_{_{\rm matt}}\right],
\end{equation}
 where
\begin{equation}\label{induced-source}
\sqrt{-g}\left(T_{\mu\nu}^{^{(D+1)}}+T^{^{[\rm
SB]}}_{\alpha\beta}\right)\equiv 2\delta\left( \sqrt{-g}\,
L\!^{^{(D)}}_{_{\rm matt}}\right)/\delta g^{\alpha\beta},
 \end{equation}
  and the induced energy momentum tensor is covariantly conserved,
namely, ${\cal D}_\beta T^{^{[\rm SB]}}_\alpha{}^{\beta}=0$.
\item
In contrast to the IMT framework where the tensor $P^{^\beta}_{\alpha}$ is
exactly conserved \cite{WO13}, the quantity $P^{^\beta}{}_{\alpha;\beta}$ in
our herein framework, in general, does not vanish, unless we focus our attention to a particular case where $\phi={\rm constant}$
 and $T^{^{(5)}}_{\alpha 4}=0$. Moreover, by assuming the extra coordinate $x^D\equiv y$ as a
cyclic coordinate, Eq.~(\ref{P-Dynamic}) reduces to an identity.
\end{enumerate}}\end{remark*}

\section{S\'{a}ez-Ballester cosmological solutions in $(D+1)$-Dimensional space-time}
\label{Bulk-solutions}
\indent
Let us start with an extended version of a spatially
flat FLRW metric in a $(D+1)$-dimensional vacuum space-time.
First, we will obtain exact cosmological solutions, and then, in the next sections,
by employing the MSBT framework, we will analyze the solutions
corresponding to the cosmology on a hypersurface. Let us consider
\begin{equation}\label{DO-metric}
dS^{2}=-dt^{2}+a^{2}(t)\left[\sum^{D-1}_{i=1}
\left(dx^{i}\right)^{2}\right]+\epsilon \psi^2(t)dy^{2},
\end{equation}
where $t$ is the cosmic time and $x^i$ (where $i=1,2,...,D-1$) denote the
Cartesian coordinates. Due to the space-time
symmetries, let us assume that the scale factor $a$ and well as the scalar fields $\phi$ and $\psi$ depend on the
comoving time only.

In the absence of ordinary matter in a $(D+1)$-dimensional
space-time, using Eqs.~(\ref{(D+1)-equation-1}) and (\ref{(D+1)-equation-2})
 for the metric~(\ref{DO-metric}), it is straightforward to show that the equations of motion are given by
\begin{eqnarray}
\label{ohanlon-eq-2}
\frac{\dot{a}}{a}\left[\frac{D-2}{2}\left(\frac{\dot{a}}{a}\right)+\frac{\dot{\psi}}{\psi}\right]=\frac{{\cal W}}{2(D-1)}\left(\phi^{\frac{n}{2}}\dot{\phi}\right)^2,\\\nonumber\\
\label{ohanlon-eq-3}
(D-2)\left[\frac{\ddot{a}}{a}+\frac{D-3}{2}\left(\frac{\dot{a}}{a}\right)^2+
\frac{\dot{a}}{a}\frac{\dot{\psi}}{\psi}\right]
+\frac{\ddot{\psi}}{\psi}=-\frac{{\cal W}}{2}\left(\phi^{\frac{n}{2}}\dot{\phi}\right)^2,\\\nonumber\\
\label{ohanlon-eq-4}
\frac{\ddot{a}}{a}+\frac{D-2}{2}\left(\frac{\dot{a}}{a}\right)^2
=-\frac{{\cal W}}{2(D-1)}\left(\phi^{\frac{n}{2}}\dot{\phi}\right)^2,\\\nonumber\\
\label{ohanlon-eq-1}
\frac{\ddot{\phi}}{\phi}+\left[(D-1)\frac{\dot{a}}{a}
+\frac{n}{2}\left(\frac{\dot{\phi}}
{\phi}\right)+\frac{\dot{\psi}}{\psi}\right]\frac{\dot{\phi}}{\phi}=0,
\end{eqnarray}
where an overdot stands for the derivative
with respect to the cosmic time $t$.

Before obtaining the general analytic solutions for the
set of Eqs~(\ref{ohanlon-eq-2})-(\ref{ohanlon-eq-1}), let us
consider the particular case where $\phi={\rm constant}$.
In this case,  Eq.~(\ref{ohanlon-eq-1}) reduces to an
identity as $0=0$, and Eq.~(\ref{ohanlon-eq-4}) yields $a(t)\propto t^{2/D}$.
Inserting this value to Eq.~(\ref{ohanlon-eq-2}), we obtain $\psi(t)\propto t^{(2-D)/D}$. In summary, we have
\begin{equation}\label{DO-metric-GR}
dS^{2}=-dt^{2}+A^{2}(t^{\frac{4}{D}})\left[\sum^{D-1}_{i=1}
\left(dx^{i}\right)^{2}\right]+\epsilon B^2t^{\frac{4}{D}-2}dy^{2},
\end{equation}
where $A$ and $B$ are constants of integration. In fact, Eq.~(\ref{DO-metric-GR}) is the
unique solution associated with the $(D+1)$-dimensional
spatially flat FLRW cosmological model in the context of
 general relativity in the absence of the ordinary matter and cosmological constant.

As it will be shown in what follows, in order to obtain exact analytic
solutions for the coupled non-linear field
Eqs.~(\ref{ohanlon-eq-2})-(\ref{ohanlon-eq-1}), we will neither
impose any ansatz nor use any simplifying condition.
Indeed, we have three unknown quantities $a(t)$, $\phi(t)$
and $\psi(t)$, which can either be related to each other or be written in terms of the cosmic time.

Using Eqs.~(\ref{ohanlon-eq-2}), (\ref{ohanlon-eq-3}) and (\ref{ohanlon-eq-1}) we obtain two constants of motion:
\begin{eqnarray}\label{new1}
a^{D-1}\phi^{\frac{n}{2}}\dot{\phi}\psi=c_1,\\\nonumber\\
\label{34}
a^{D-1}\dot{\psi}=c_2,
\end{eqnarray}
where $c_1$ and $c_2$ are constants of integration. Equations (\ref{new1}) and (\ref{34}) yield an exact solution by
which $\psi$ can be written in terms of $\phi$ for different values of $n$:
\begin{equation}\label{new7}
\psi=\left \{
 \begin{array}{c}
  \psi_i {\rm Exp}\left(\frac{2\beta}{n+2}\phi^{\frac{n+2}{2}}\right)
 \hspace{15mm} {\rm for}\hspace{5mm} n\neq-2,\\\\
 \psi_i\phi^\beta
  \hspace{33mm} {\rm for}\hspace{5mm} n=-2,
 \end{array}\right.
\end{equation}
where $\psi_i$ is an integration constant, $\beta\equiv\frac{c_2}{c_1}$ such
that we have assumed $c_1\neq0$ and $a^{D-1}\psi\neq0$.
Substituting $\psi$ from Eq.~(\ref{new7}) into Eq.~(\ref{ohanlon-eq-2}) we obtain a relation between $a$ and $\phi$, which also depends on $n$
\begin{equation}\label{new17}
a=\left \{
 \begin{array}{c}
  a_i {\rm Exp}\left[\frac{2\gamma(D)}{n+2}\phi^{\frac{n+2}{2}}\right]
 \hspace{18mm} {\rm for}\hspace{5mm} n\neq-2,\\\\
 a_i\phi^{\gamma(D)}
  \hspace{33mm} {\rm for}\hspace{5mm} n=-2,
 \end{array}\right.
\end{equation}
where $a_i$ is a constant of integration and parameter $\gamma(D)$ is defined as\footnote{Obviously, $\gamma$
is also a function of $\beta$ and ${\cal W}$, but we have denoted it as $\gamma(D)$ to
 emphasise that it depends on the number of dimensions.}
\begin{eqnarray}\label{new17-2}
\gamma(D)\equiv\frac{1}{D-2}\left[-\beta\pm\sqrt{\beta^2+\left(\frac{D-2}{D-1}\right){\cal W}}\right].
\end{eqnarray}
Since $\gamma(D)$ should be a real parameter, therefore the allowed range for ${\cal W}$ is
\begin{eqnarray}\label{W-range}
{\cal W} \geq -\left(\frac{D-1}{D-2}\right)\beta^2,
\end{eqnarray}
which implies that ${\cal W}$ can take positive or negative values.

Now, we can obtain three unknowns $a$, $\phi$, and $\psi$ in terms of the cosmic time.
Let us proceed as follows. Substituting $\psi$ and $a$, respectively,
from Eqs.~(\ref{new7}) and (\ref{new17}) into differential Eq.~(\ref{new1}) we obtain
\begin{equation}\label{new18-2}
\left \{
 \begin{array}{c}
 \dot{\phi}\phi^{\frac{n}{2}}{\rm Exp}\left[\frac{2f(D)}{n+2}\phi^{\frac{n+2}{2}}\right]=\frac{c_1a_i^{1-D}}{\psi_i}
 \hspace{12mm} {\rm for}\hspace{5mm} n\neq-2,\\\\\\
   \dot{\phi}\phi^{f(D)}=\frac{c_1a_i^{1-D}}{\psi_i}
  \hspace{30mm} {\rm for}\hspace{5mm} n=-2,
 \end{array}\right.
\end{equation}
where, for convenience, we defined $f(D)$ as
\begin{eqnarray}\label{f}
f(D)\equiv (D-1)\gamma(D)+\beta.
\end{eqnarray}
Equations (\ref{new18-2}) yield two classes of exact
solutions in terms of the cosmic time, which
can be categorised in terms of different values taken by the quantity $f(D)$.
In this respect, let us proceed our discussions as follows.

\subsection{Case~I:\,$f(D)\equiv(D-1)\gamma+\beta=0$}
Eq.~(\ref{new17-2}) implies
\begin{eqnarray}\label{Gamma-W}
\gamma=-\frac{\beta}{D-1}, \hspace{10mm} {\cal W}=-\left(\frac{D}{D-1}\right)\beta^2;
\end{eqnarray}
note that ${\cal W}$ is in the allowed range indicated by (\ref{W-range}).

Moreover, Eq.~(\ref{new18-2}) yields
\begin{equation}\label{new54}
 \phi(t)=\left \{
 \begin{array}{c}
\left[\frac{(n+2)(1-D)h(D)(t-t_i)}{2 \beta}\right]^{\frac{2}{n+2}}
 \hspace{12mm} {\rm for}\hspace{8mm} n\neq-2,\\\\
  {\rm Exp}\left[\frac{(1-D)h(D)(t-t_i)}{\beta}\right]
  \hspace{17mm} {\rm for}\hspace{8mm} n=-2,
 \end{array}\right.
\end{equation}
where $t_i$ is an integration constant and
\begin{equation}\label{H-d}
h(D)\equiv \frac{c_1\beta a_i^{1-D}}{(1-D)\psi_i}.
\end{equation}
Therefore, by substituting
$\phi(t)$ from Eq.~(\ref{new54}) to Eqs.~(\ref{new7}), (\ref{new17}),
the scale factor $a$ and the scalar field $\psi$ can be re-written in terms of the cosmic time as
\begin{eqnarray}\label{new59}
 a(t)\!\!&=&\!\!a_i\,{\rm Exp}\left[h(D)\left(t-t_i\right)\right] ,\hspace{19mm} \forall n\\\nonumber\\
 \label{new59-2}
 \psi(t)\!\!&=&\!\!\psi_i\,{\rm Exp}\left[(1-D) h(D)(t-t_i)\right],\hspace{8mm} \forall n
\end{eqnarray}
which implies that (i) when the $a$ and $\psi$ are expressed in terms of the
cosmic time, they do not depend explicitly on $n$;
(ii) for $ n\neq-2$, the scalar field $\phi(t)$ is a power-law
function of the cosmic time, but $a(t)$ and $\psi(t)$ are exponential
functions of the cosmic time for all values of $n$.

Using Eqs.~(\ref{new59}) and (\ref{new59-2}), metric (\ref{DO-metric}) can be written as
\begin{equation}\label{CaseI-metric}
dS^{2}=-dt^{2}+a_i^2{\rm Exp}\left[2 h(D)(t-t_i\right)]\left[\sum^{D-1}_{i=1}
\left(dx^{i}\right)^{2}\right]+\epsilon \psi_i^2{\rm Exp}\left[2(1-D) h(D)(t-t_i)\right]dy^{2}.
\end{equation}
On a $D$-dimensional hypersurface\footnote{The reduced cosmology will be
studied comprehensively in the next section.}, solutions (\ref{new54}) and (\ref{CaseI-metric})
associated with $n=-2$ bear much resemblance to the de Sitter-like
solution obtained in the context of the Brans-Dicke (BD) theory for a spatially flat FLRW vacuum universe \cite{RFM14}.
For instance, assuming the special case where $D=4$ and $\beta=1$, our herein model corresponds to a four-dimensional
vacuum spatially flat FLRW model obtained in the context of the BD theory
for the particular value of the BD coupling parameter, $\omega_{_{\rm BD}}=-4/3$ \cite{Faraoni.book} such
that $\phi$ in our herein model plays the role of the BD scalar field.
However, as it will be shown in the
 next section, in our herein MSBT model, in contrast
 to the mentioned BD cosmological model, the $D$-dimensional
 hypersurface is not empty. Concretely, such a generalised solution emerges not only from an
 induced matter but also in the presence of self interacting scalar potential.
 Moreover, by assuming $h(D)>0$, both the scalar field and the
 extra dimension always decrease.


\subsection{Case~II$:f(D)\equiv(D-1)\gamma(D)+\beta\neq0$}

Let us first express ${\cal W}$ in terms of
other parameters. Eq.~(\ref{new17-2}) reads
 \begin{eqnarray}\label{w-ii}
 {\cal W}=(D-1)\gamma(D) [2\beta+(D-2)\gamma(D)].
 \end{eqnarray}
Integrating both sides of Eq.~(\ref{new18-2}) over $dt$ yields
\begin{equation}\label{new20-22}
 \phi(t)=\left \{
 \begin{array}{c}
\left\{\frac{n+2}{2f(D)}{\rm ln}\left[\tilde{h}(D)(t-t_i)\right]\right\}^{\frac{2}{n+2}}
 \hspace{16mm} {\rm for}\hspace{8mm} n\neq-2,\\\\
  \left[\tilde{h}(D)(t-t_i)\right]^{\frac{1}{f(D)}}
  \hspace{28mm} {\rm for}\hspace{8mm} n=-2,
   \end{array}\right.
\end{equation}
where
\begin{equation}\label{h-tild}
\tilde{h}(D) \equiv \frac{c_1f(D)}{a_i^{D-1}\psi_i}.
\end{equation}
In order to obtain the scale factors in terms of the cosmic time, we substitute Eq.~(\ref{new20-22})
into Eqs.~(\ref{new7}) and (\ref{new17}) and get
\begin{eqnarray}\label{new29}
 a(t)\!\!&=\!\!&a_i \left[\tilde{h}(D)(t-t_i)\right]^{r},\hspace{8mm} \forall n\\\nonumber
 \\\label{new29-2}
   \psi(t)\!\!&=&\!\! \psi_i \left[\tilde{h}(D)(t-t_i)\right]^{m},\hspace{8mm} \forall n,
 \end{eqnarray}
 where, as $f(D)\neq0$, for the later convenience, we introduced new parameters as
 \begin{eqnarray}\label{r-m}
r\equiv\frac{\gamma}{f(D)},\hspace{8mm} m\equiv\frac{\beta}{f(D)},\hspace{8mm} {\rm where} \hspace{8mm} m+(D-1)r=1.
 \end{eqnarray}
Obviously, $r$ and $m$ depend also on the number of dimensions.
Moreover, with these definitions, relation (\ref{w-ii}) can be rewritten as
\begin{eqnarray}\label{w-ii-2}
 {\cal W}=(D-1)f^2(D)r [2m+(D-2)r].
 \end{eqnarray}

Again, let us focus on a $D$-dimensional hypersurface. For $n=-2$, as
both $\phi(t)$ and $a(t)$ are power-law forms in terms of the
cosmic time, and $1/f+(D-1)m=1$, therefore this case correspond to the generalised version of the
O'Hanlon-Tupper solution \cite{RFM14} when the BD coupling parameter
is restricted as $\omega_{_{\rm BD}}>-(D-1)/(D-2)$ and $\omega_{_{\rm BD}}\neq -D/(D-1)$.
However, similar to the exponential-law
solutions, and in contrary with the O'Hanlon-Tupper model, the
induced energy momentum tensor does not generally vanish on the hypersurface.
We conclude that these resemblance between our herein
solutions for the particular case where $n=-2$ and those obtained within the BD theory
might represent the correspondence between the Einstein frame and Jordan frame.


\section{Reduced Cosmology on a $D$-dimensional hypersurface}
\label{Red-cosm}
\indent
Here using the MSBT framework established in Section \S \ref{Set up} and
employing the exact cosmological solutions obtained in the previous section,
we will present reduced cosmological
dynamics on a $D$-dimensional hypersurface.
We will focus on the particular characteristic features of the
solutions when they provide conditions which yield not only an
 accelerating scale factor $a(t)$ but also a simultaneous contracting scalar $\psi(t)$.

Since $a(t)$, $\psi(t)$ and $\phi(t)$ are independent of the
extra coordinate $y$, from Eqs.~(\ref{DO-metric}) and (\ref{matt.def}) we can show that
the components of the induced energy momentum tensor
on a hypersurface $\Sigma_0$ are:
\begin{eqnarray}\label{R16}
\rho_{_{\rm SB}}\equiv - T^{0[{\rm SB}]}_{\,\,\,0}\!\!\!&=&\!\!\!
\frac{1}{\chi}\left[\frac{\ddot{\psi}}{\psi}-\frac{V(\phi)}{2}\right],\\\nonumber
\\
\label{R17}
p_{_{\rm SB}}\equiv T^{i[{\rm SB}]}_{\,\,\,i}\!\!\!&=&\!\!\!
-\frac{1}{\chi}\left[\frac{\dot{a}\dot{\psi}}{a\psi}-\frac{V(\phi)}{2}\right],
\end{eqnarray}
where $i=1,2,3, ... ,D-1$ (with no sum);
$\rho_{_{\rm SB}}$ and $p_{_{\rm SB}}$ denote the induced energy
density and pressure, respectively. Moreover, to compute the
induced scalar potential, we should use Eq.~(\ref{v-def}), leading to
\begin{equation}\label{new30}
\frac{dV}{d\phi}{\Biggr|}_{_{\Sigma_{o}}}\!\!\!\!\!=2{\cal W}\phi^n\dot{\phi}\left(\frac{\dot{\psi}}{\psi}\right).
\end{equation}
Now, considering the constants of motion given by Eqs.~(\ref{new1}) and (\ref{34}),
$\dot{\phi}$ and $\dot{\psi}$ can be removed in favour of the other variables of the model:
\begin{equation}\label{new30-1}
\frac{dV}{d\phi}{\Biggr|}_{_{\Sigma_{o}}}\!\!\!\!\!=2c_1^2\beta{\cal W}a^{2(1-D)}\phi^{\frac{n}{2}}\psi^{-2}.
\end{equation}
As we will see in the following,  there are two
different classes of solutions. More precisely,
substituting $a$ and $\psi$ from Eqs.~(\ref{new7})
and (\ref{new17}) to Eq.~(\ref{new30}), yields two different equations as
\begin{equation}\label{new31}
\frac{dV}{d\phi}{\Biggr|}_{_{\Sigma_{o}}}=\left \{
 \begin{array}{c}
 V_0\phi^{-\left[1+2f(D)\right]}
 \hspace{25mm} {\rm for}\hspace{8mm} n=-2,\\\\
   V_0\phi^{\frac{n}{2}}{\rm Exp}\left[-\frac{4f(D)}{n+2}\phi^{\frac{n+2}{2}}\right]
  \hspace{12mm} {\rm for}\hspace{8mm} n\neq-2,
 \end{array}\right.
\end{equation}
where
\begin{equation}\label{V0}
V_0\equiv 2c_1^2\beta{\cal W}a_i^{2(1-D)}\psi_i^{-2}.
 \end{equation}
We will proceed the calculations associated with the induced
matter and scalar potential in the following subsections.

Equations (\ref{BD-Eq-DD}) and \eqref{D2-phi} for the induced metric (i.e., the $D$-dimensional
spatially flat FLRW metric) can be written as
\begin{eqnarray}
\label{DD-FRW-eq1}
\frac{(D-1)(D-2)}{2}H^2\!\!&=&\!\!\chi\rho_{_{\rm SB}}+\rho_{\phi}\equiv \rho_{_{\rm tot}},\\\nonumber\\
\label{DD-FRW-eq2}
(D-2)\frac{\ddot{a}}{a}+\frac{(D-2)(D-3)}{2}H^2
\!\!&=&\!\!-\left(\chi p_{_{\rm SB}}+p_{\phi}\right)\equiv -p_{_{\rm tot}},\\\nonumber\\
\label{DD-FRW-wave}
2\phi^n\ddot{\phi}+2(D-1)H\phi^n\dot{\phi} &+&n\phi^{n-1}\dot{\phi}^2+\frac{1}{{\cal W}}\frac{dV}{d\phi}{\Biggr|}_{_{\Sigma_{o}}}=0.
\end{eqnarray}
Equations (\ref{DD-FRW-eq1}) and  (\ref{DD-FRW-eq2}) lead to
\begin{eqnarray}
\label{DD-FRW-eq3}
\frac{\ddot{a}}{a}
=-\frac{1}{(D-1)(D-2)}\left[(D-3)\rho_{_{\rm tot}}+(D-1)p_{_{\rm tot}}\right].
\end{eqnarray}
In Eqs.~(\ref{DD-FRW-eq1})-(\ref{DD-FRW-wave}),
$H\equiv \dot{a}/a$ is the Hubble parameter and $\rho_{_{\rm SB}}$,
$p_{_{\rm SB}}$  and $\frac{dV}{d\phi}{\Big|}_{_{\Sigma_{o}}}$ are given by
Eqs.~(\ref{R16}), (\ref{R17}) and (\ref{new30}) respectively.
Moreover, $\rho_{\phi}$ and $p_{\phi}$ are
the energy density and pressure associated with the scalar field $\phi$:
\begin{eqnarray}
\label{rho-phi-gen}
\rho_\phi\!\!&\equiv\!\!&\frac{1}{2}\left[{\cal W}
\phi^n\dot{\phi}^2+V(\phi)\right],
\\\nonumber\\
\label{p-phi-gen}
p_\phi\!\!&\equiv\!\!&\frac{1}{2}\left[{\cal W}
\phi^n\dot{\phi}^2-V(\phi)\right].
\end{eqnarray}
Let us now introduce the following quantities:\\
 EoS parameters are defined as
\begin{eqnarray}
\label{eos}
W_{_{\rm SB}}\equiv \frac{p_{_{\rm SB}}}{\rho_{_{\rm SB}}}, \hspace{10mm}
W_\phi\equiv \frac{p_\phi}{\rho_\phi}, \hspace{10mm}
 W_{_{\rm tot}}\equiv \frac{p_{_{\rm tot}}}{\rho_{_{\rm tot}}}=\frac{\chi p_{_{\rm SB}}+p_{\phi}}{\chi\rho_{_{\rm SB}}+\rho_{\phi}}.
\end{eqnarray}
We may also use the deceleration parameter $q=-a\ddot{a}/(\dot{a})^2$.
Assuming the FLRW metric and
conservation of the energy momentum tensor, the present values of $H$ and $q$ are
independent of the applied gravitational theory \cite{SB85-original}.
\\
We define the density
parameters associated with the induced matter and the scalar field, respectively, as
\begin{eqnarray}
\label{density.par.def1}
\Omega_{_{\rm SB}}&\equiv&\frac{2\chi}{(D-1)(D-2)}\frac{\rho_{_{\rm SB}}}{H^2},\\\nonumber\\
\label{density.par.def2}
\Omega_{\phi}&\equiv&\frac{2}{(D-1)(D-2)}\frac{\rho_{\phi}}{H^2}.
\end{eqnarray}
Consequently, Eq.~(\ref{DD-FRW-eq1}) can be written as $\Omega_{_{\rm SB}}+\Omega_{\phi}=1$.

In what follows, as both the energy density and pressure associated
to the scalar field take positive and negative values,
let us review different types of matter, see for instance \cite{NJP15}.
Suppose a general case whose EoS and density parameters are defined as $W\equiv p/\rho$
and $\Omega\equiv \rho/\rho_c$, respectively (where $\rho_c\equiv (D-1)H^2/8\pi G>0$
is the critical density in a $D$-dimensional space-time). (i) For a positive energy
density (i.e., $\rho>0$), there are two different cases with
negative pressure ($p<0$) and positive
pressure ($p>0$), which are called a dark energy and light energy, respectively \cite{NJP15}.
(ii) Conversely, for a negative energy density, the dark energy and light energy correspond to the
cases whose pressure is positive and negative, respectively.
 Definitely, the density parameter corresponding to
 cases (i) and (ii) is positive and negative, respectively.
  As an interesting solution obtained from our
  herein model, we will show that, under suitable conditions
 among the parameters of the model, it is feasible to get the total
 energy density as either a positive cosmological
 constant or a quintessence (with positive energy density).
 More concretely, both of these total energy densities are comprising from
 a positive energy density $\rho_{_{\rm SB}}$, which is coupled
 with a negative energy density $\rho_{\phi}$.

From Eqs.~(\ref{DD-FRW-eq1}), (\ref{DD-FRW-eq2}),
and Eqs.~(\ref{rho-phi-gen}) and (\ref{p-phi-gen}), one concludes
that there are similarities between the field equations of theory obtained
 from Einstein-Hilbert action including a canonical scalar
 field (which responds to the potential energy $\tilde{V}(\phi))$ and our model in the particular
 case where $n=0$, ${\cal W}=1$, $\tilde{V}(\phi)\equiv 1/2V(\phi)$.
 However, we should note that there is a noticeable distinction
 between these frameworks: in our herein model, not only the components of the induced matter
 (i.e., $p_{_{\rm SB}}$ and $\rho_{_{\rm SB}}$) but also the
 potential energy of the $\rho_{\phi}$ and $\rho_{\phi}$
 emerge from the geometry of
 the extra dimension, see Eq.~(\ref{v-def}).
 Such modifications are the distinguishing features of the
 IMT, MBDT and MSBT, which not only may have the same properties of the
 ordinary matter in the universe, but also can play the role of dark matter and
 dark energy within suitable circumstances.

Whether or not the
quantity $f(D)$ vanishes, each equation of (\ref{new31}), in turn, will yield
different solutions. Namely, we will show that there are two different
classes of solutions, which will be investigated in separated cases as follows.

\subsection{Case~I:\,$f(D)=(D-1)\gamma(D)+\beta=0$}
\indent \label{CaseI}
For this case, integrating Eq.~(\ref{new31}) yields
\begin{equation}\label{new66}
V(\phi)=\left \{
 \begin{array}{c}
\frac{2V_0}{n+2}\,\phi^{\frac{n+2}{2}}
 \hspace{17mm} {\rm for}\hspace{8mm} n\neq-2,\\\\
 V_0\,{\rm ln}\left(\frac{\phi}{\phi_i}\right)  \hspace{15mm} {\rm for}\hspace{8mm} n=-2,
 \end{array}\right.
\end{equation}
where $\phi_i$ is an integration constant and substituting ${\cal W}$ from
Eq.~(\ref{Gamma-W}) into Eq.~(\ref{V0}), and using Eq.~(\ref{H-d}), $V_0$ for this case can be rewritten as
\begin{equation}\label{V0-caseI}
V_0=2\beta D(1-D)h^2(D).
\end{equation}
In order to derive the induced scalar potential in terms of the cosmic time, we substitute
Eq.~ (\ref{new54}) into Eq.~(\ref{new66}). Therefore, we obtain a
unique relation of the potential energy for all $n$:
\begin{equation}\label{new69}
V(t)=V_0(1-D)h(D)\beta^{-1}\,(t-t_i)=2D(1-D)^2h^3(D)\,(t-t_i), \hspace{8mm} \forall n,
\end{equation}
in which $h(D)$ is given by Eq.~(\ref{H-d}).

We can now obtain the components of the induced matter on a $D$-dimensional hypersurface.
By substituting the induced scalar potential (\ref{new69}), the scale
factors and their time derivatives from Eqs.~(\ref{new59}), (\ref{new59-2})
(\ref{R16}) and (\ref{R17}), we get
\begin{eqnarray}\label{new74}
\chi\rho_{_{\rm SB}}\!\!&\equiv\!\!&-\chi T^{0[{\rm SB}]}_{\,\,\,0}=\left(1-D\right)^2h^2(D)
\left[-Dh(D)(t-t_i)+1\right],\\\nonumber
\\\label{new75}
\chi p_{_{\rm SB}}\!\!&\equiv\!\!&\chi T^{j[{\rm SB}]}_{\,\,\,j}=\left(1-D\right)h^2(D)
\left[D\left(1-D\right)h(D)(t-t_i)-1\right],
\end{eqnarray}
where $j=1,2,...,D-1$.
 As mentioned, conservation of the induced energy momentum
 tensor in MSBT is one of the characteristic features of our model.
 Concerning this case, using Eqs.~(\ref{new59}), (\ref{new74}) and (\ref{new75}), we
 can show that the
 quantity $\dot{\rho}_{_{\rm SB}}+(D-1)H(\rho_{_{\rm SB}}+p_{_{\rm SB}})$ identically vanishes.

Substituting the scalar field and the induced scalar
potential from Eqs.~(\ref{new54}) and (\ref{new69}) into Eqs.~(\ref{rho-phi-gen})
and (\ref{p-phi-gen})), we obtain the energy density and the pressure associated with the scalar field:
\begin{eqnarray}\label{ro-phi-deSit}
\rho_{\phi}&=&\frac{D(1-D)h^2(D)}{2}\left[1+2\left(1-D\right)h(D)(t-t_i)\right], \hspace{5mm} \forall n,\\\nonumber\\
\label{p-phi-deSit}
p_{\phi}&=&\frac{D(1-D)h^2(D)}{2}\left[1-2\left(1-D\right)h(D)(t-t_i)\right], \hspace{5mm} \forall n,
\end{eqnarray}
leading to
\begin{eqnarray}
\label{w-phi-deSit}
W_{\phi}=\frac{1-2\left(D-1\right)h(D)(t-t_i)}{1+2\left(D-1\right)h(D)(t-t_i)},\hspace{5mm} \forall n.
\end{eqnarray}
It is worth to summarise the solutions for this case:
\begin{eqnarray}\label{CaseI-metric-hyper}
ds^{2}&=&-dt^{2}+a_i^2{\rm Exp}\left[2 \sqrt{\frac{\Lambda(D)}{D-1}}\left(t-t_i\right)\right]\left[\sum^{D-1}_{i=1}
\left(dx^{i}\right)^{2}\right],\\\nonumber
\\\label{CaseI-metric-hyper-2}
\psi(t)&=&\psi_i^2{\rm Exp}\left[2(1-D)\sqrt{\frac{\Lambda(D)}{D-1}}(t-t_i)\right],\\\nonumber\\
\label{w-tot-deSit}
\rho_{_{\rm tot}}&=&\frac{(D-2)}{2}\Lambda(D),\hspace{5mm}
p_{_{\rm tot}}=-\frac{(D-2)}{2}\Lambda(D), \hspace{5mm}
W_{\rm tot}=-1,\hspace{5mm}   \forall n,
\end{eqnarray}
where
\begin{eqnarray}
\label{lambda-deSit}
\Lambda(D)\equiv (D-1)h^2(D)=\frac{1}{D-1}\left(\frac{c_1\beta}{a_i^{D-1}\psi_i}\right)^2={\rm constant}> 0, \hspace{5mm} {\rm for} \hspace{3mm} D>1.
\end{eqnarray}
To obtain the total energy density and
pressure, we substitute $\rho_{_{\rm SB}}$, $p_{_{\rm SB}}$, $\rho_{\phi}$, and $p_{\phi}$
from Eqs.~(\ref{new74}), (\ref{new75}), (\ref{ro-phi-deSit}) and (\ref{p-phi-deSit})
into  Eqs.~(\ref{DD-FRW-eq1}) and (\ref{DD-FRW-eq2}).

Due to the presence of the positive cosmological
constant, which emerges from two varying matter
fields, (\ref{new74})-(\ref{p-phi-deSit}), we conclude that
we have a exponentially expanding universe, which corresponds to the de Sitter solution.

Let
us now look at the allowed values taken by $\Lambda(D)$.
Using the definition $\beta\equiv\frac{c_2}{c_1}$ and the constant of motion
(\ref{p-phi-deSit}), Eq.~(\ref{lambda-deSit}) can be written as
\begin{eqnarray}
\label{lambda-deSit-2}
\Lambda(D)=\frac{1}{D-1}\left(\frac{\dot{\psi}}{\psi}\right)^2={\rm constant},\hspace{5mm} \forall t,
\end{eqnarray}
which is satisfied at any time.

More concretely, Eq.~(\ref{lambda-deSit}) indicates
that the cosmological constant depends only on the number
of space-time dimensions, integration constants as
well as parameters of model; or equivalently, as
Eq.~(\ref{lambda-deSit-2}) implies, it is at at any specific time directly
obtained from the values taken by the squared
expansion rate associated with the extra
dimension at that time over the number of spatial dimensions.
Furthermore, it is a positive constant.



\subsection{Case~II:\,$f(D)\equiv\beta+(D-1)\gamma(D)\neq0$}
\indent \label{CaseII}
In this case, Eq.~(\ref{new31}) yields
the induced scalar potential as
\begin{equation}\label{new32}
V(\phi)=\left \{
 \begin{array}{c}
-\frac{V_0}{2f(D)}\,{\rm Exp}\left[\frac{-4f(D)}{n+2}\,\phi^{\frac{n+2}{2}}\right]
\hspace{15mm} {\rm for}\hspace{8mm} n\neq-2,\\\\
 -\frac{V_0}{2f(D)}\phi^{-2f(D)}
\hspace{30mm} {\rm for}\hspace{8mm} n=-2.
 \end{array}\right.
\end{equation}
Substituting the scalar field from Eq.~(\ref{new20-22}),
the induced potential is written in terms of the cosmic time:
\begin{equation}\label{new40}
V(t)=-\frac{(D-1)mr \left[1+m-r\right]}{(t-t_i)^{2}}, \hspace{10mm} \forall n,
\end{equation}
which, similar to the case I, it is independent of $n$.

Combining Eqs.~(\ref{new29}), (\ref{new29-2}) and (\ref{new40})
with Eqs.~(\ref{R16}) and (\ref{R17}), one finds
\begin{eqnarray}\label{new46}
\rho_{_{\rm SB}}=-\frac{D(D-1)m r^2}{2\chi (t-t_i)^{2}},
\hspace{15mm} p_{_{\rm SB}}=-\frac{D m r\left(1+m\right)}{2\chi (t-t_i)^{2}},  \hspace{10mm} \forall n.
\end{eqnarray}
These relations imply that (i) to demand $\rho_{_{\rm SB}}\geq0$, $m$ should
take negative values, which, in turn, indicates that the extra
dimension decreases with cosmic time; (ii) to get non-negative values
for $\rho_{_{\rm SB}}+p_{_{\rm SB}}=-Dmr/\chi (t-t_i)^{2}$, it is necessary to assume $r>0$ and $m<0$.
Concretely, (i) and (ii) imply that in order to satisfy the weak energy condition
for the induced matter, $a(t)$ and $\psi(t)$ should increase and decrease with cosmic time, respectively.

Equations~(\ref{new46}) lead to
\begin{eqnarray}\label{new49}
p_{_{\rm SB}}=W_{_{\rm SB}}\rho_{_{\rm SB}}, \hspace{10mm} {\rm where} \hspace{10mm}
W_{_{\rm SB}}=\frac{1+m}{(D-1)r},  \hspace{10mm} \forall n,
\end{eqnarray}
which implies that the induced matter on the brane obeys the
EoS of a barotropic fluid. Moreover, it is independent of $n$.

Substituting $a(t)$, $\rho_{_{\rm SB}}$ and $p_{_{\rm SB}}$ from
Eqs.~ (\ref{new17}) and (\ref{new46}) into
$\dot{\rho}_{_{\rm SB}}+3H(\rho_{_{\rm SB}}+p_{_{\rm SB}})$, one finds that it identically vanishes; namely, the induced matter is conserved.

Let us also obtain the energy density and the pressure associated
with the scalar field. Substituting $\phi(t)$ and $V(t)$ from Eqs.~(\ref{new20-22}) and (\ref{new40})
 into Eqs.~(\ref{rho-phi-gen}) and (\ref{p-phi-gen}), for all values of $n$, we obtain:
 \begin{eqnarray}\label{kin-phi}
{\cal W}\phi^n\dot{\phi}^2&=&\frac{ (D-1)r(1+m-r)}{2 (t-t_i)^2},\\\nonumber\\
\label{ro-phi}
\rho_{\phi}&=&\frac{{\cal W}(1-m)}{2f^2(t-t_i)^2}=\frac{\left[(D-1)r\right]^2
\left[2m+(D-2)r\right]}{2(t-t_i)^2},\\\nonumber\\\nonumber\\
\label{p-phi}
p_{\phi}&=&\frac{{\cal W}(1+m)}{2f^2(t-t_i)^2}=\frac{\left[(D-1)r\right]
\left[2m+(D-1)r\right]\left[2m+(D-2)r\right]}{2(t-t_i)^2},
\end{eqnarray}
which gives
\begin{eqnarray}
\label{w-phi}
W_{\phi}&=&\frac{1+m}{1-m}=\frac{(D-1)r+2m}{(D-1)r},\\\nonumber\\\nonumber\\
\label{omega-phi}
\Omega_{\phi}&=&\frac{(1-m)(1+m-r)}{(D-2)r}=
\frac{(D-1)[(D-2)r+2m]}{(D-2)[(D-1)r+m]}.
\end{eqnarray}

Substituting $\rho_{_{\rm SB}}$, $p_{_{\rm SB}}$, $\rho_{\phi}$, and $p_{\phi}$
from Eqs.~(\ref{new46}), (\ref{ro-phi}) and (\ref{p-phi})
 into definitions (\ref{DD-FRW-eq1}) and (\ref{DD-FRW-eq2}), the total energy density and pressure are given by
\begin{eqnarray}
\label{w-tot}
\rho_{_{\rm tot}}=\frac{(D-1)(D-2)r^2}{2(t-t_i)^2},\hspace{5mm}
p_{_{\rm tot}}=\frac{(D-2)r\left(1+m\right)}{2(t-t_i)^2}, \hspace{5mm}
W_{\rm tot}=\frac{1+m}{(D-1)r}\hspace{5mm}   \forall n.
\end{eqnarray}
Up to now, we have shown that the total matter on the hypersurface is a
barotropic matter which, in turn, is obtained from adding two other barotropic matter fluids.
It is easy to show that the following conservation equations are satisfied:
 \begin{eqnarray}
\label{con-tot-phi}
\dot{\rho}_{\phi}+3H(\rho_{\phi}+p_{\phi})=0, \hspace{8mm}
\dot{\rho}_{_{\rm tot}}+3H(\rho_{_{\rm tot}}+p_{_{\rm tot}})=0.
\end{eqnarray}
Moreover, in this
case (where $f\neq0$ and for all $n$), we get $W_{\rm tot}=W_{_{\rm SB}}=W_{\phi}$.
According to the definitions of the EoS parameters, such a result can
 also be obtained in approximation conditions
 where $|p_{\phi}|\ll \chi|\rho_{_{\rm SB}}| $ and/or $\chi |p_{_{\rm SB}}|\ll |\rho_{\phi}|$.

Finally, substituting $\rho_{_{\rm tot}}$ and $p_{_{\rm tot}}$ from Eq.~(\ref{w-tot}) into Eq.~(\ref{DD-FRW-eq3}), we obtain

\begin{eqnarray}
\label{acc-fin}
  \frac{\ddot{a}}{a}=-\frac{r\left[m+(D-2)r\right]}
 {(t-t_i)^2},\hspace{5mm} \forall n.
\end{eqnarray}
Hence, to get an accelerating
scale factor, the following inequality should always be satisfied
\begin{eqnarray}\label{acc-con2}
r\left[m+(D-2)r\right]<0, \hspace{8mm} {\rm or} \hspace{8mm}\gamma\left[\beta+(D-2)\gamma\right]<0,\hspace{5mm} \forall n.
\end{eqnarray}
The above requirement for an expanding universe (i.e., $r\equiv \gamma/f>0$), leads to obtain constraints
among the parameters of the model, which will be discussed in two separated cases as follows.
Before focussing on details, without loss of generality, let us
assume the following conditions for the constant coefficients
 in the scale factors in Eqs.~(\ref{new29}) and (\ref{new29-2})
\begin{eqnarray}\label{acc-con3}
a_i \tilde{h}^r>0, \hspace{8mm} \psi_i \tilde{h}^m>0.
\end{eqnarray}
Moreover, as the following considerations are valid for all $n$,
let us refrain from writing $\forall n$ in front of equations.
Furthermore, we will assume $D>2$ and let us just focus on the signs
of the quantities, without obtaining their exact ranges.

\begin{description}
  \item[Case IIa) $\gamma<0$, $\beta+(D-2)\gamma>0$:] In this case, it is easy to show that
  \begin{eqnarray}\label{acc-con3}
-\frac{\beta}{D-2}<\gamma<0, \hspace{7mm} \beta>0, \hspace{7mm} \gamma<f,\hspace{7mm}
 2(D-1)\beta\gamma<{\cal W}<(D-1)\beta\gamma<0.
\end{eqnarray}
Moreover, $f(D)$ is restricted to $-\beta/(D-2)<f(D)<\beta$, namely, it
can take positive and negative values.
However, for an expanding universe, we should have $\gamma/f>0$ for the power-law solution (\ref{new29}). Therefore,
from Eqs.~(\ref{new29}), (\ref{acc-con2}) and (\ref{acc-con3}), we conclude
that, $f(D)$ must take negative values. Consequently, the following range is allowed for $f(D)$:
\begin{eqnarray}\label{acc-con3}
-\frac{\beta}{D-2}<f(D)<0.
\end{eqnarray}
Up to now, we have shown that, by admitting the above conditions upon the parameters of the model,
 $a>0$, and $\ddot{a}>0$; whilst $\psi(t)$ decreases with cosmic time, which seems
desirable in the context of Kaluza-Klein frameworks~\cite{OW97}.

Let us now proceed our investigation by obtaining the allowed
ranges for the energy density, pressure as well as density parameters.
Admitting Eqs.~(\ref{acc-con2})-(\ref{acc-con3}), we have shown that, at
any arbitrary time, the following inequalities are valid
\begin{eqnarray}
\label{acc-con4-1}
\rho_{_{\rm SB}}&>&0, \hspace{8mm} p_{_{\rm SB}}<0, \hspace{8mm} W_{_{\rm SB}}<0, \hspace{8mm} \Omega_{_{\rm SB}}>0, \\\nonumber\\
\label{acc-con4-2}
\rho_{\phi}&<&0, \hspace{8mm}p_{\phi}>0, \hspace{9mm} W_{\phi}<0, \hspace{9mm} \Omega_{ \phi}<0, \\\nonumber\\
\label{acc-con4-3}
\rho_{_{\rm tot}}&>&0, \hspace{8mm}p_{_{\rm tot}}<0, \hspace{8mm} W_{_{\rm tot}}<0,
\end{eqnarray}
where we have used $f+\beta>0$ (that is satisfied for $D>3$).
Moreover, we should note that, for the case IIa, both the potential ($V(t)$) and
the kinetic (${\cal W}\phi^n\dot{\phi}^2$) contributions of the energy
density associated with the scalar field $\phi$ take negative values at all times.

  \item[Case IIb) $\gamma>0$, $\beta+(D-2)\gamma<0$:] It is straightforward to show that
   \begin{eqnarray}\label{acc-con5}
0<\gamma<-\frac{\beta}{D-2}, \hspace{7mm} \beta<0, \hspace{7mm} \gamma>f, \hspace{7mm}  2(D-1)\beta\gamma<{\cal W}<(D-1)\beta\gamma<0.
\end{eqnarray}

Again, we have shown that $f(D)$ can take both positive and
negative values from the interval $\beta<f(D)<-\beta/(D-2)$.
However, similar to the Case IIa, again we should take $\gamma/f>0$.
Concretely, in this case, $f$ must take positive values. Namely, we obtain
 \begin{eqnarray}\label{acc-con6}
0<f(D)<-\frac{\beta}{(D-2)}.
\end{eqnarray}

Consequently, by admitting the above constraints on the parameters of the
model and assuming $D>3$, we have shown that the
inequalities (\ref{acc-con4-1})-(\ref{acc-con4-3}) are again satisfied
for the physical quantities of this case.
\end{description}

Let us summarise the SB solutions associated with this case from the view an
observer who lives on the $D$-dimensional space-time and
is not aware of the existence of the extra dimension.
Therefore, we should remove the parameter $m$ in favour of the others. We obtain
\begin{eqnarray}\label{DO-metric-final}
ds^2&=&dS^2{\Big|}_{_{\Sigma_{y}}}=-dt^{2}+a_i^2
\left[\tilde{h}(D)(t-t_i)\right]^{2r}\left[\sum^{D-1}_{i=1}
\left(dx^{i}\right)^{2}\right],\\\nonumber\\
\label{rho-final}
\rho_{_{\rm SB}}&=&-\frac{D(D-1)[1-(D-1)r]r^2}{2\chi (t-t_i)^{2}},
\hspace{10mm}p_{_{\rm SB}}=W_{_{\rm SB}}\rho_{_{\rm SB}},\\\nonumber\\
V(t)&=&-\frac{(D-1) r(2-Dr)[1-(D-1)r]}{(t-t_i)^{2}},
\end{eqnarray}
and the SB scalar field $\phi(t)$ is given by (\ref{new20-22}).
Using Eqs.~(\ref{acc-con4-3}), $W_{_{\rm SB}}$
we can rewrite (\ref{acc-con4-3}) in terms of $r$ only:
\begin{eqnarray}\label{DO-metric-final}
W_{_{\rm SB}}=\frac{2}{(D-1)r}-1.
\end{eqnarray}
Using Eqs.~ (\ref{r-m}) and (\ref{w-ii-2}) we get:
\begin{eqnarray}\label{w-ii-3}
 {\cal W}=\frac{\left(D-1\right)r \beta^2 \left(2-Dr\right)}{\left[1-\left(D-1\right)r\right]^2},
 \end{eqnarray}
 where $\beta$ is the ratio of the constants of motion.

 For the power-law solution,to get an accelerating
 universe, we should have $q=-(r-1)/r<0$, i.e., $r>1$.
  Therefore, all the three EoS parameters must be less than $(3-D)/(D-1)$. (For instance, assuming a
  five-dimensional bulk, we get $W_{_{\rm SB}}(=W_{_{\rm tot}}=W_{\phi})<-1/3$.)
  Namely, assuming $D>3$, at all times, the energy density associated with both the induced
  matter and the total matter take positive values, whilst whose pressures take negative values.
  More concretely, both of these matters play the role of the dark energy.
  Moreover, we still obtain $(\rho_{_{\rm SB}}
+p_{_{\rm SB}}), (\rho_{_{\rm tot}}+p_{_{\rm tot}})\geq 0$.
Namely, the weak energy condition for both of these kinds of matter is satisfied.

 Concerning the matter associated with $\phi$, using $r>1$ in Eq.~(\ref{w-ii-3}), we find that ${\cal W}<0$.
 Namely, to get an accelerating universe for the
  power-law solution, from Eqs.~(\ref{kin-phi})-(\ref{omega-phi}), we obtain
${\cal W}\phi^n\dot{\phi}^2<0$, $\rho_{\phi}<0$, $p_{\phi}>0$ and $\Omega_{\phi}<0$.
 Such kind of matter can be considered as a dark energy \cite{NJP15}.
 In addition, in this case, $m$ takes negative values.
 Therefore, for an accelerating universe, the extra dimension shrinks with cosmic time.

For an ordinary matter, the corresponding EoS parameter $W$ should be restricted as $0\leq W \leq 1$
(which, specially, includes the matter-dominated, radiation-dominated and stiff
fluid with $W=0, 1/(D-1), 1$ in a $D$-dimensional space-time).
Therefore,  Eq.~(\ref{DO-metric-final}) yields $1/(D-1)\leq r \leq 2/(D-1)$, which
for $D>3$ corresponds to a
decelerating universe. Consequently, we can easily show
that $\rho_{_{\rm SB}}, \rho_{_{\rm tot}} \geq 0$ and $(\rho_{_{\rm SB}}
+p_{_{\rm SB}}), (\rho_{_{\rm tot}}+p_{_{\rm tot}})\geq 0$.
Namely, the weak energy condition is satisfied for both the induced
and total matters. For the above range of $r$ we obtain $-1\leq m \leq 0$, which
implies that the extra dimension shrinks with cosmic time.
However, concerning the matter associated with the SB scalar field,
according to Eqs.~(\ref{kin-phi})-(\ref{p-phi}), it is
straightforward to show ${\cal W}$, $\rho_\phi$ and $p_\phi$ take
negative and positive values in the ranges $2/D < r \leq 2/(D-1)$
and $1/(D-1)\leq r <2/D$, respectively, and they
vanish when $r=2/D$. These ranges also determine
whether or not the weak energy condition is satisfied for this matter.

\begin{figure}
\centering\includegraphics[width=2.6in]{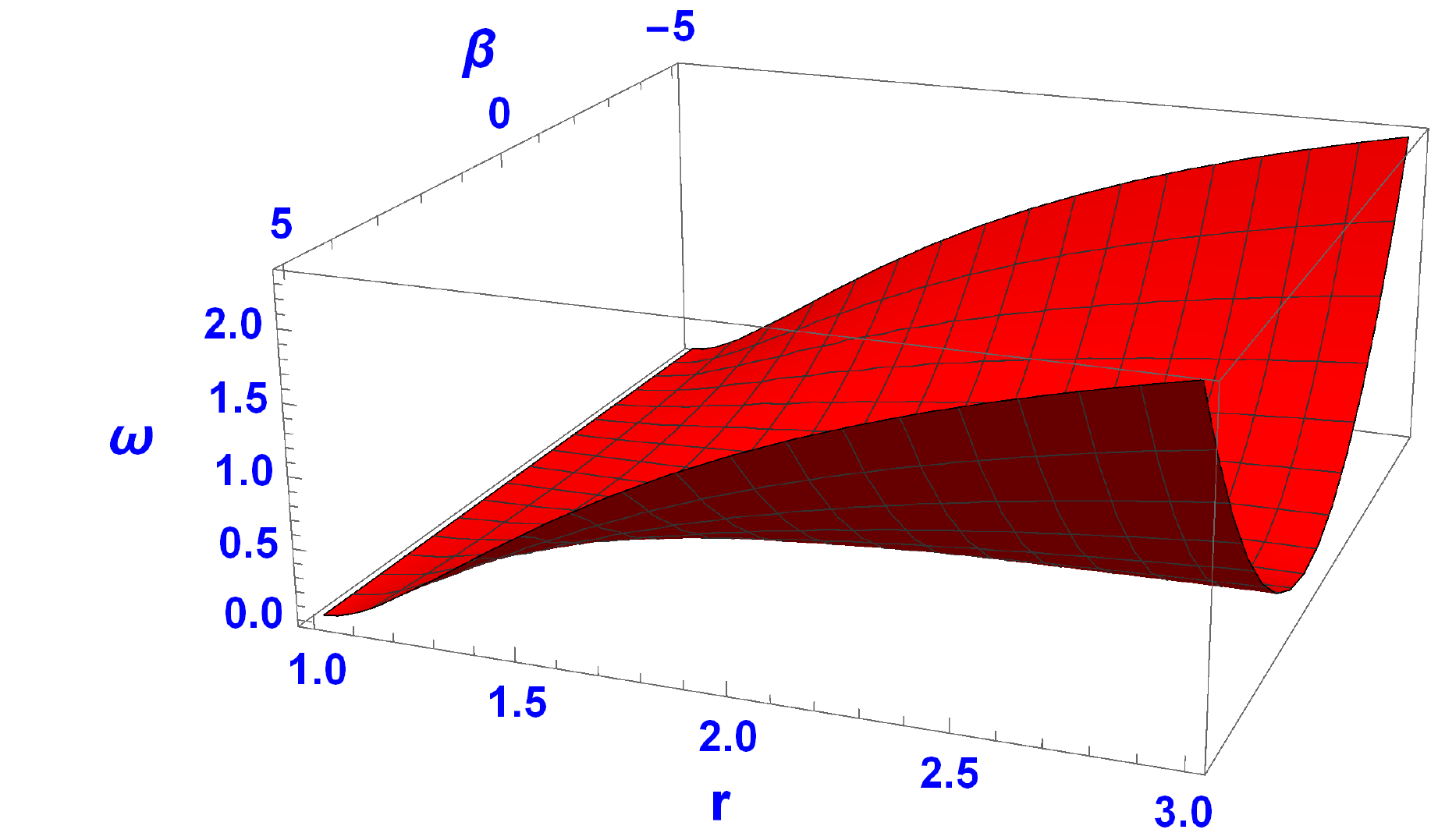}
\centering\includegraphics[width=2.6in]{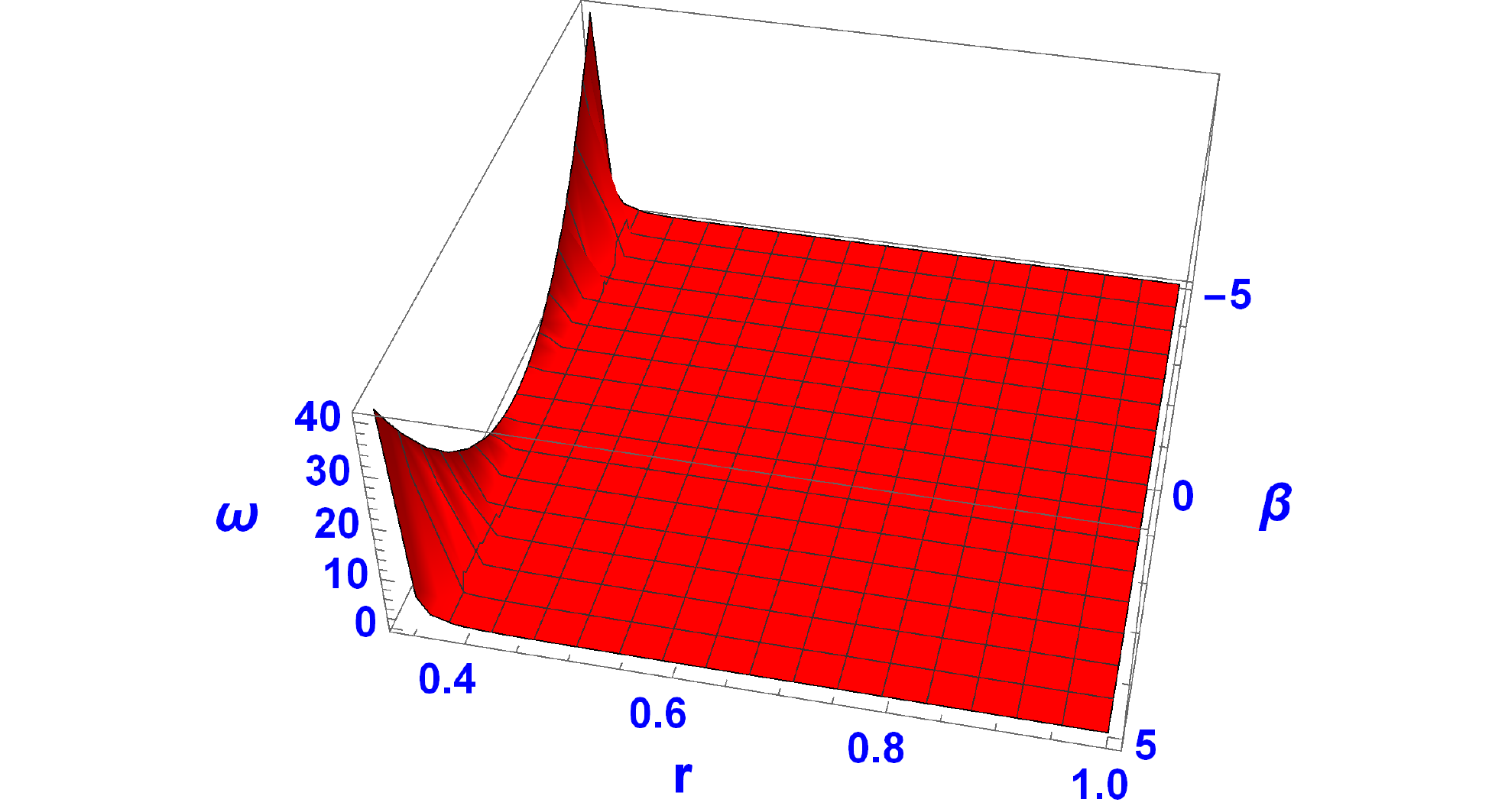}
\caption{The behaviour of $\omega$ in the allowed region of the
parameter space $(r, \beta)$ for both accelerating and
decelerating phases for a particular case where $D=4$.
Because of satisfaction of the weak energy condition, we have assumed $r>1/3$.
}
\label{omega}
\end{figure}

As it was shown for accelerating and decelerating scale factors, ${\cal W}$ takes
negative and positive values, respectively. It is worthwhile to check also the satisfaction of the
constraint \eqref{W-range}. For more clarity of the plots, let us define instead
\begin{eqnarray}
\label{omega-range}
  \omega\equiv {\cal W} +\left(\frac{D-1}{D-2}\right)\beta^2\geq 0.
\end{eqnarray}
For $D=4$, figures \ref{omega} show that the constraint
\eqref{omega-range} (or equivalently \eqref{W-range}) is satisfied
appropriately for both accelerating and decelerating phases.

\section{Summary and Discussion}
\indent \label{conclusion}
In Section \S \ref{Set up}, by employing a generalised procedure, we have
established an extended version of the MSBT with an induced matter
and a self-interacting scalar potential in arbitrary dimensional hypersurface.
 In particular, the SB field equations were derived from a generalised $(D+1)$-dimensional action.
In this action, neither scalar potential nor cosmological constant
have been considered. However, in order to extend the
procedure used in the original IMT, we have assumed the bulk to contain ordinary matter.

Using a general and coordinate free framework, the geometrical quantities
associated with the $(D+1)$-dimensional semi-Riemannian
manifold have been related to those corresponding to a $D$-dimensional one.
These general equations have been applied
 to the particular metric \eqref{global-metric}.

Using the latter relations for the SB field equations, we
have established the $D$-dimensional SB field
equations, in which the induced matter and the scalar potential
have an intrinsic geometrical origin. One of the characteristic features of
this model is that the reduction procedure generates the type of the induced potential.

We should emphasise that, due to the
 coordinate free framework we have used, our procedure to
 obtain the MSBT is much broader than those used to
 construct the standard IMT \cite{PW92}, MBDT \cite{RFM14} or MSBT \cite{RM18}.

As a cosmological application of the MSBT, we considered a
universe that is described with an extended version of an
FLRW metric in a $(D+1)$-dimensional space-time.
By assuming a simple case in which there is no ordinary
matter, we have shown that there are two constants of motion
constructed from different functions of the three unknowns of the model.
Then, we have obtained two distinct classes of exact solutions for the field equations.
We have expressed the scale factors
and the scalar field in terms of the cosmic time, such that for
both of the classes, the scale factors do not
explicitly depend on the parameter $n$ (see action \eqref{SB-5action}), whilst, the scalar field $\phi(t)$,
depends on $n$, and it is given by two different functions of the cosmic time.
Moreover, we have demonstrated how these solutions on the hypersurface
correspond to the solutions obtained in the BD cosmology for
the spatially flat FLRW universe; cf Section \S \ref{Bulk-solutions}.

Subsequently,
we have focused our attention to the cosmological solutions on
the hypersurface.
Accordingly, we have introduced useful physical quantities to analyze our solutions.
Moreover, in order to clarify the reduced cosmology, we have also established the
standard Friedmann equations (which are obtained for the
spatially flat FLRW metric within the context of general relativity)
on the hypersurface, whose (total) matter field in the right
 hand sides is composed of two non-interacting components: the induced matter (which
 directly emerges from the geometry of the extra
 dimension) and the matter associated with the SB scalar field $\phi$
 (whose potential, according to our herein MSBT, has also geometrical origin).

 As mentioned earlier, there are two distinct classes of solutions on the hypersurface.
 In what follows, in addition to presenting a summary of our two classes, we provide a complementary discussion.

\begin{description}
  \item[Case I) Cosmological constant:]

  For the first class of solutions, we have found that a positive
cosmological constant dominates, which is, surprisingly, obtained
from superposition of two time dependent sectors.
Consequently, the scale
factor of the universe expands exponentially, which can be used to describe either inflation at the early
universe or the acceleration at late times. However, the scalar
field $\psi(t)$ decreases with comic
time, which indicates that the extra dimension contracts or shrinks.
Note that the so-called cosmological
constant appearing in the effective field equations is
not added by hand; it
depends on the number of dimensions of the space time and the expansion rate of the
scalar field $\psi(t)$ (i.e., the scale factor associated with the non-compact extra dimension).

In the solution associated with the spatially flat FLRW metric (\ref{DO-metric}) within
the context of IMT, the scale factor of the universe cannot ever accelerate.
 As the unique solution (\ref{DO-metric-GR}) indicates, in
the context of the IMT, neither the
cosmological constant nor an accelerating phase is
obtained for the universe described with the FLRW metric (\ref{DO-metric}).
 In order to convey a better understanding of the nature of the cosmological
constant, the canonical metric has been employed in noncompact KK
gravity, see, e.g., \cite{MLW94,MWL98} and related papers.
For instance, on a $D$-dimensional hypersurface, we can find a correspondence between our asymptotic
solution (i.e., the solution obtained in the particular case where $f(D)=0$) and
that obtained in the IMT for the canonical metric
\begin{eqnarray}\nonumber
dS^{2}=\frac{y^2}{L^2}\tilde{g}_{\mu\nu}(x^\alpha, y)dx^\mu dx^\nu-dy^2,
\end{eqnarray}
where $L>0$ is a constant.

Let us consider a particular case where $T_{\mu\nu}^{^{(D+1)}}=0$, ${\tilde{g}}_{\mu\nu}={\rm diag}(1, -a^2(t),
-a^2(t), ..., -a^2(t))$ and $\phi=\phi(t)$. For this case, the induced matter (\ref{T1}) is given by
\begin{eqnarray}\nonumber\label{can-2}
\chi T_{\mu\nu}^{^{(D)[{\rm eff}]}}=\frac{1}{2L^2}
\left(D^2-3D+2+y_0^2V_0\right){\tilde{g}}_{\mu\nu},
\end{eqnarray}
where $y_0={\rm constant}$ is the value of $y$ on the hypersurface and
 the constant potential $V_0$ is obtained from (\ref{v-def}).
Therefore, Eq.~(\ref{BD-Eq-DD}) reduces to
\begin{eqnarray}\nonumber\label{can-1}
G_{\mu\nu}^{^{(D)}}=\left(\frac{D^2-3D+2}{2L^2}\right){\tilde{g}}_{\mu\nu}
+{\cal W}\phi^n\left[({\cal D}_\mu\phi)({\cal D}_\nu\phi)-\frac{y_0^2}{2L^2}({\cal D}_\alpha\phi)({\cal
D}^\alpha\phi){\tilde{g}}_{\mu\nu}\right].
\end{eqnarray}

Then the SB cosmological
solutions associated with the canonical gauge are given by
\begin{eqnarray}\nonumber\label{can-3}
a(t)&=&a_i {\rm Exp}\left[\frac{(D-1)(t-t_i)}{L}\right],\\\nonumber\\\nonumber
 \phi(t)&=&\left \{
 \begin{array}{c}
\Bigg\{\frac{CL(n+2)}{2(1-D)}{\rm Exp}\left[\frac{(1-D)(t-t_i)}{L}\right]\Bigg\}^{\frac{2}{n+2}}
 \hspace{12mm} {\rm for}\hspace{8mm} n\neq-2,\\\\
  \Bigg\{{\rm Exp}\left[{\rm Exp}(\frac{(1-D)(t-t_i)}{L})\right]\Bigg\}^{\frac{CL}{1-D}}
  \hspace{17mm} {\rm for}\hspace{8mm} n=-2,
 \end{array}\right.,
\end{eqnarray}
where $a_i$ and $t_i$ are integration constants and $C$ is the constant of motion given by
$C\equiv a^{D-1}\phi^{\frac{n}{2}}\dot{\phi}$.
Setting $\phi={\rm constant}$, we can easily show that our SB model reduces to the
corresponding one obtained within the context of IMT,
i.e., $G_{\mu\nu}=(D^2-3D+2){\tilde{g}}_{\mu\nu}/(2L^2)$, in
which $\Lambda=\Lambda(D)\equiv(D^2-3D+2)/(2L^2)$; for the case
 where $D=4$, see, for instance, \cite{MWL98,P08} and references therein.

 \item[Case II) Power-law solutions:]
The solutions of this case, in terms of the cosmic time, for
both the $a(t)$, $\psi(t)$ and the induced scalar potential $V(t)$ are
in power-law forms; while the SB scalar field $\phi(t)$ is in power-law and
logarithmic forms for $n=-2$ and $n\neq-2$, respectively. We have shown that
 both the induced matter and the matter associated with the scalar
 field satisfy the barotropic EoS; they also obey the conservation law.
 Namely, the total matter in this case is composed of two non-interacting fluids.
 We have obtained the corresponding density parameters.

 Subsequently, we have focused on the behaviour of the quantities for
 different values of the present parameters and integration constants.
We have shown that it is feasible to get an accelerating universe.
 In this case, the total matter, the
  induced matter and the matter associated
  with $\phi$ could play the role of a dark energy.
  However, in the case of a decelerating universe, all three
  fluids play a role as ordinary matter in the universe for the
  allowed ranges of the parameters.
  We have discussed how to satisfy the weak energy condition for all kinds of matter.
  Furthermore, we have shown that for both the accelerating and the decelerating
  expansions, the large extra dimension contracts with cosmic time.
\end{description}

In what follows, let us further add some brief points.

\begin{itemize}

  \item
We have shown that the MSBT is described by four equations of
motion (\ref{D2say}), (\ref{BD-Eq-DD})(\ref{D2-phi}) and (\ref{P-Dynamic}).
Concerning the cosmological example, which is described with the
extended FLRW metric (\ref{DO-metric}), in section \ref{Red-cosm}, we
just used Eqs.~(\ref{BD-Eq-DD})-(\ref{D2-phi}).
As there is no ordinary matter in the bulk and
the coefficient of the metric (\ref{DO-metric}) as well as the
scalar field depend on the cosmic time only (imposing the cylinder condition), therefore,
(i) from Eq.~(\ref{D2say}), we obtain ${\cal D}^2\psi=0$, which
is the Klein-Gordon equation for the massless scalar field.
Moreover, $T^{^{(D)[{\rm eff}]}}=g^{\mu\nu}T_{\mu\nu}^{^{(D)[{\rm eff}]}}=V(\phi)/2\chi$, which indicates
that, contrary to the IMT, the induced matter in the MSBT, even
by imposing the cylinder condition and assuming a bulk in the
absence of the ordinary matter, does necessarily consist of photons only.
It is seen that  Eq.~(\ref{D2say}), using the solutions (\ref{new59}), (\ref{new59-2}), (\ref{new29}), (\ref{new29-2}), is
identically satisfied for all the values taken by $n$ and $f(D)$.
(ii) Eq. (\ref{P-Dynamic}) implies that $P_{\alpha\beta}$ is a conserved quantity.

\item
As in the particular cases where $n=-2$ and $n=0$, the SB framework transforms
 to the well-known canonical scalar field model
  (for $n=-2$, we can set $\phi\equiv Exp \left(\Phi\right)$), to demand non-ghost
 scalar fields, we should exclude the negative values of ${\cal W}$.
 However, for general cases where $n\neq -2,0$, determining whether or
  not the scalar field being a ghost field seems not only depend on the values of ${\cal W}$, but also
  the values of $n$ and the behavior of $\phi(t)$ at all times, which must be inspected carefully.

  \item
  To the best of our knowledge, the observational constraints on
  the parameters of the SB theory have not been studied.
  Such objectives were however beyond the scope of this work.

\end{itemize}

\section*{Acknowledgments}
We would like to thank Prof. James M. Overduin and Arash Ranjbar for their fruitful comments.
SMMR appreciates the support of the grant SFRH/BPD/82479/2011 by
the Portuguese Agency Funda\c{c}\~ao para a Ci\^encia e
Tecnologia. The work of M.S. is supported in part by the Science and Technology Facility Council (STFC), United Kingdom, under the research grant ST/P000258/1. PVM is grateful to DAMTP and Clare Hall, Cambridge for
kind hospitallity and a Visiting Fellowship during his sabbatical. This research work was supported by
Grant No. UID/MAT/00212/2019 and COST Action CA15117 (CANTATA).

\bibliographystyle{utphys}

\end{document}